\documentclass[twocolumn,prc,preprintnumbers,nofootinbib,superscriptaddress,showpacs,a4]{revtex4-1}
\usepackage{epsfig,graphicx,float,pst-all}
\usepackage{mathrsfs}
\usepackage[utf8]{inputenc}
\usepackage{rotating}
\usepackage{url}
\usepackage{esvect}
\usepackage{subfigure}
\usepackage{multirow}
\usepackage{amsmath}
\usepackage{amssymb}
\usepackage{xcolor}
\usepackage{natbib}
\usepackage{hyperref}
\usepackage[warn]{textcomp}
\usepackage{gensymb}
\usepackage{footnote}
\usepackage{siunitx}
\usepackage{comment}
\usepackage{bm}
\usepackage{orcidlink}
\usepackage[normalem]{ulem}

\newcommand{\nucl}[2]{^{#1}\mathrm{#2}}

\begin{document}
%\title{Mechanism of blitz/snap inversion in $pf$-shell leading to halos in Na Isotopes }
\title{The effect of inversion of $p$ and $f$ orbits on halo formation in heavy sodium isotopes}
\author{Jagjit Singh\orcidlink{0000-0002-3198-4829}}
\email{Jagjit.Singh@manchester.ac.uk}
\affiliation {Department of Physics and Astronomy, The University of Manchester, Manchester M13 9PL, UK}
\affiliation{Department of Physics, Akal University, Talwandi Sabo, Bathinda, Punjab 151302, India}
\affiliation{Research Center for Nuclear Physics (RCNP), Osaka University, Ibaraki 567-0047, Japan}
\author{J. Casal\orcidlink{0000-0002-5997-5860}}
\email{jcasal@us.es}
\affiliation{Departamento de F\'{i}sica At\'{o}mica, Molecular y Nuclear, Facultad de F\'{i}sica, Universidad de Sevilla, Apartado 1065, E-41080 Sevilla, Spain}
\author{L. Fortunato\orcidlink{0000-0003-2137-635X}}
\email{lorenzo.fortunato@pd.infn.it}
\affiliation{Dipartimento di Fisica e Astronomia ``G.Galilei'', Università degli Studi di Padova, via Marzolo 8, Padova, I-35131, Italy}
\affiliation{INFN-Sezione di Padova, via Marzolo 8, Padova, I-35131, Italy}
\author{N. R. Walet\orcidlink{0000-0002-2061-5534}}
\email{Niels.Walet@manchester.ac.uk}
\affiliation {Department of Physics and Astronomy, The University of Manchester, Manchester M13 9PL, UK}
\date{\today}
\begin{abstract}
The role of the inversion of the $p$  and $f$ shell-model orbits in the emergence of halo structures in the ground states of neutron-rich $\nucl{34,37,39}{Na}$ is investigated. Families of two- and three-body models are constructed with effective core-neutron interactions, with parameter choices based on a combination of the available experimental data and systematic trends, as well as the GPT $n$-$n$ interaction and a phenomenological three-body force. Our results indicate a possible one-neutron halo in $\nucl{34}{Na}$, while $\nucl{37,39}{Na}$ exhibit features of Borromean halos. The halo formation is driven by the weakening of the shell gap and inversion of the $2p_{3/2}$ and $1f_{7/2}$ orbits expected to occur somewhere near these masses. We further show that the electric dipole response provides a clear and sensitive probe of halo structure in these isotopes.
\end{abstract}
\maketitle

\section{Introduction}

When moving away from the valley of stability, where standard stable nuclei are located, we will eventually reach the drip lines, where nuclei become unstable against emitting nucleons. Here, one encounters short-lived isotopes with extreme neutron-to-proton ratios. Such exotic nuclei exhibit unique properties and behaviours that are very different from those of stable nuclei, and thus of substantial interest \cite{Filomena2021,Crawford2024,Brown2025_FRIB}. Since it is challenging to produce and study such nuclei experimentally, measured data can be scarce or absent. If we venture into this uncharted territory, we need guidance, which requires informed speculation about the nature of the terrain. 

This is precisely the case of neutron-rich sodium isotopes, with masses ranging from $A=32$ to $A=39$ ($N=21$ to $N=28$). According to the traditional single-particle shell model the neutrons should fill the $f_{7/2}$ shell. When the neutron number increases, nuclei become progressively more weakly bound. This results in weakening and broadening of the single-particle potential, which weakens the gap between the $f$ and $p$ orbitals and eventually leads to a degeneracy or inversion of the normal ordering. This effect is well-known in elements with slightly lower proton numbers \cite{Hammamoto2007,Hammamoto2012,Kobayashi2016,Kahlbow2024}.
 
\begin{figure*}[!ht]
\centering
\includegraphics[width=0.9\linewidth]{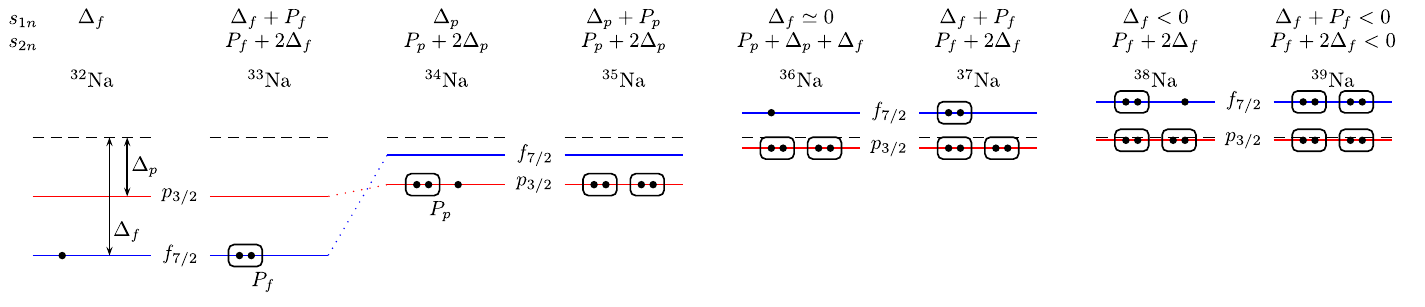}
\caption{Schematic representation of a potential scenario for the evolution of neutron shell-model orbits in the independent-particle shell model and the associated change in filling for sodium isotopes with $A={32, \cdots, 39}$ (energies not to scale). $\Delta_\ell$ (arrows) and $P_\ell$ (boxes) represent the single-particle energy level gaps to the threshold and correlation energies, respectively. 
An inversion between $p$ and $f$ orbits occurs between $^{33}$Na and $^{34}$Na, then the $f$ orbit gets very close to threshold at $A=36$ and becomes definitely unbound for $A=38$.} 
\label{Fig1}
\end{figure*}	
	
Let us now have a quick systematic look at what the data tells us about the heavy sodium nuclei. We show a simple schematic illustration of our analysis in Fig.~\ref{Fig1}. Here we model the one- and two-neutron separation energies as $s_{1n}=P_\ell+\Delta_\ell$ for even $N$ and $s_{1n}=\Delta_\ell$ for odd $N$, and  $s_{2n} = P_\ell +2\Delta_\ell$, where $\Delta_\ell$ is the orbital energy with respect to threshold and $P_\ell$ the energy required to break a pair. One can then use the available evaluated \cite{Wang21} information on $s_{1n}$ and $s_{2n}$ to determine the energies of the valence orbit with respect to the separation threshold (see Fig.~\ref{Fig1} and  Table \ref{TL1}). 

%32Na
Following the normal shell model ordering, we expect that $^{32}_{11}$Na$_{21}$ has one neutron in the $1f_{7/2}$ orbit, and that seems in agreement with experiment. 
%33Na
For $^{33}_{11}$Na$_{22}$, which has a larger one neutron separation energy, $s_{1n}$, due to the effect of pairing for an even number of neutrons, the standard ordering of the neutron shells may still hold. The contribution of the odd proton in the $d_{5/2}$ shell is split over  a number of states due to a substantial deformation, as suggested by $\gamma$-ray spectroscopy experiments on the $K=3/2$ ground state band and Monte-Carlo shell model calculations in Ref.~\cite{Gade2011}. 

%34Na
The weak binding in $^{34}_{11}$Na$_{23}$ allows for the possibility that it is a one-neutron-halo nucleus. Experimental support comes from direct mass measurements \cite{Gaudfroy2012} and a preliminary interaction cross-section analysis \cite{Kuboki2011,Suzuki2014}. Theoretical investigations \cite{Fortune2013,GSingh2016,GSingh2017,Manju19EPJ,Luo2020} agree with this speculation. Thus, $^{34}_{11}$Na$_{23}$ may represent a turning point in the evolution of nuclear structure in heavy sodium isotopes. % See Fig.~\ref{fig:inversion}. 
%35Na
The isotope $^{35}_{11}$Na$_{24}$ is more deeply bound than its predecessor, and thus unlikely to be a halo nucleus.

%36Na
The next nucleus, $^{36}_{11}$Na$_{25}$, has not been observed, and is expected to be marginally unbound. If this is the case, our simple way of determining the single-particle energies should be modified: Since the one neutron separation energy is compatible with zero,  the energy required to lift two neutrons is $s_{2n}= P_p+\Delta_p+\Delta_f \approx P_p+\Delta_p$. This means that in Table \ref{TL1}, in this case, we should not evaluate the pairing energy, but rather the sum  $P_p+\Delta_p$, indicated by an asterisk.
%37Na
Continuing on, $^{37}_{11}$Na$_{26}$ is barely bound again, suggesting a potential two-neutron halo with significant mixing of these energetically close shells. The nature of this state depends on details of the shell structure. If the $p_{3/2}$ state is lowest and fully occupied by bound neutrons, the two valence neutrons in $^{37}_{11}$Na$_{26}$ would occupy the $f_{7/2}$ level, which is not favourable for halo formation. 
In a situation in which both levels have similar energies, close to the Fermi energy, we expect a smooth occupancy and some interesting halo physics will appear due to the significant percentage of $p$-wave character.  

%38Na
The next odd-$N$ isotope, $^{38}_{11}$Na$_{27}$, is one-neutron unbound. 
%39Na
Even though the evaluation \cite{Wang21}  suggests that $^{39}_{11}$Na$_{28}$ is unbound as well, this nucleus has been observed \cite{Ahn2022}, and found to be bound--as yet there is no estimate for the separation energy. This is an excellent candidate for a Borromean halo. Studies using relativistic Hartree-Bogoliubov theory suggest that the $N=28$ shell gap in $^{39}$Na is significantly reduced \cite{Zhang2023}. This is compatible with our analysis: the pairing energy is about the same value as the gap, which keeps both orbitals close to threshold.

\begin{table}[tbh]
\caption{Estimate of the energy gaps to threshold, $\Delta$, and pairing energy $P$, see Fig.~\ref{Fig1} and discussion in the main text. All values are in MeV.}
\label{TL1}
\centering
\begin{tabular}{ccccccccc}
\\[-1ex]
%%%%%%%
\toprule\\[-1.2ex]
A&32&33&34&35&36&37&38&39 \\[1ex]
%%%%%%%%
\toprule\\[-1.2ex]
$\Delta$ &1.68&1.68&0.17&0.17&0.&0.&-0.7&-0.7 \\[1ex]
%\toprule
%MeV&&&&&&&&\\
$P$ &-&1.25&2.76&1.75&1.92*&0.84&1.54&0.7 \\[1ex]
%%%%%%%%%%%%%%%%%%%%%%%%%%%%%%%%%%%%%%%%%%%%%%%%%%%%%%%%%%%%%%%%%%%%%%%%%%%%%%%%%%%%%%
\toprule
\end{tabular}
\end{table}

%We have tabulated our simple-minded estimates for the values of $\Delta$ and $P$ in Table \ref{TL1} with notation referring to Fig.~\ref{Fig1} and data taken from Ref.~\cite{Wang21}.

In this study, we perform a systematic horizontal scan of the neutron-rich Na isotopic chain within a few-body framework, reaching nuclei for which experimental information is extremely limited.  The few-body approach is particularly effective near the neutron drip line, where the structure is governed by one or two weakly bound valence neutrons and continuum effects become essential. This will allow us to develop scenarios that should be testable by experiment. Note that few-body approaches, while typically more precise in the description of weakly bound nuclei, rely on existing data to constrain model parameters. In the model we describe the single-particle states in the core nucleon system with a simple residual interaction, similar to the independent particle shell model, but there are important differences. One of the most important differences is that the potential for these weakly bound neutrons is much more diffuse than for the shell model, and has no deeply bound states (achieved by an appropriate transformation).  In this way, our calculations reflect much of what is seen in Fig.~\ref{Fig1}.
Using this model, we analyse the possible scenarios for a one-neutron halo in $\nucl{34}{Na}$ and also the detailed structure of the putative Borromean halos $\nucl{37}{Na}$ and $\nucl{39}{Na}$. This is followed by a detailed study of the $B(E1)$  response in all nuclei studied.
%Although the model is intentionally simple, it captures the essential physics governing these neutron-rich systems and enables meaningful predictions across the isotopic chain. 
%The present results are therefore timely and are expected to stimulate new experimental and theoretical investigations. 

\section{Model Formulation}\label{sec:MF}
We describe the effective $\text{core}+n$ and $\text{core}+2n$ systems in a discrete pseudostate representation~\cite{JALay10,JCasal13}. The corresponding two- and three-body Hamiltonians are diagonalized in a discrete basis, which provides bound states and a discrete $L^2$ approximation of the continuum states. 

For two-body systems, the single-particle calculations in a discrete basis are straightforward, provided one knows effective $\text{core}+n$ potentials. For the three-body case, wave functions are constructed using the hyperspherical harmonics framework~\cite{ZHUKOV1993,Nielsen01}. One starts from the usual Jacobi coordinates $\{\boldsymbol{x},\boldsymbol{y}\}$, which describing the relative motion between two particles and the centre of mass with respect to the third particle, respectively. For a $\text{core}+2n$ system, we focus on the Jacobi-$T$ representation, where the neutron separation is $x$, and $y$ is the distance of their centre of mass to the core. 
The states of the system are given by
\begin{equation}
  \Psi^{j\mu}(\rho,\Omega)=\frac{1}{\rho^{5/2}}\sum_{\beta}\chi_{\beta}^j(\rho) \mathcal{Y}_{\beta}^{j\mu}(\Omega),
 \label{eq:3bwf}
\end{equation}
with the hyper-radius $\rho=\sqrt{x^2+y^2}$ and the hyperangular variable $\Omega=\{\alpha,\widehat{x},\widehat{y}\}$, where $\tan\alpha=y/x$. The indices $\beta$, called the channels, represent the quantum numbers %in the chosen coupling order of angular momenta and spins 
compatible with $j$, and $\mathcal{Y}_{\beta}^{j\mu}$ are the appropriate hyperspherical harmonics. These functions depend on the hypermomentum $K$, which defines the effective three-body centrifugal barrier in the Schrödinger equation for the hyperradial coordinate. The functions $\chi_{\beta}^j(\rho)$, the hyperradial wavefunctions, are expanded in a transformed harmonic oscillator (THO) basis, which has the advantage of reproducing the correct exponential decay for bound states,
\begin{equation}
 \chi_{\beta}^j (\rho) = \sum_{i}C_{i\beta}^j U_{i\beta}^{\text{THO}}(\rho).
 \label{eq:expTHO}
\end{equation}
In practice, a finite three-body Hamiltonian matrix is built by choosing a maximum value of hypermomentum, $K\leq K_{max}$, which also sets the maximum value of the relative angular momenta in the two Jacobi subsystems (i.e., $l_x+l_y\leq K$~\cite{ZHUKOV1993}), and a total number of basis functions $N$ ($i=0,1,\dots,N$). Diagonalizing this matrix we obtain the expansion coefficients for bound and (discretized) continuum eigenstates. This Hamiltonian includes all pair-wise interactions, constrained by the available information on the corresponding binary subsystems. In addition, a phenomenological three-body force depending on the hyperradius is incorporated to account to fix the correct separation energy of the system, if experimentally known. This also takes into account some effects missing from the truncation of the model space. The reader is referred to previous works for more details~\cite{JCasal13,JCasal20}. 

\section{Two-body potentials}\label{2bpots}
In an effective description of one-neutron halo nuclei and the core-neutron subsystems of Borromean systems, the key ingredient is the core-neutron interaction; this depends on experimental data, but we need a parametrisation that is both simple and sufficiently rich. In this work, we model the interaction using a Woods-Saxon potential with central and spin-orbit terms. We vary the central strength, but the spin-orbit strength is determined from systematics following Ref.~\cite{Horiuchi201031Ne}. This procedure gives a spin-orbit strength of $28.00\,\mathrm{MeV\,fm^2}$ for $\nucl{32}{Na}$. For simplicity, we use the same value for all Na isotopes considered in the present work. Furthermore, we also use the same range parameter, $r_0=1.25\,\mathrm{fm}$, and diffuseness, $a=0.75\,\mathrm{fm}$, for the central and spin-orbit terms. 

These phenomenological $\text{core}+n$ potentials sustain also bound states in lower orbits, which are associated to Pauli-forbidden states already occupied by the core neutrons. For the three-body calculations of $^{37,39}$Na, these states need to be projected out of the available space. Here, as in Refs.~\cite{JCasal20,JSingh2020}, this is achieved by constructing phase-equivalent shallow potentials.

\begin{figure}[t]
\centering
\includegraphics[width=1.1\linewidth]{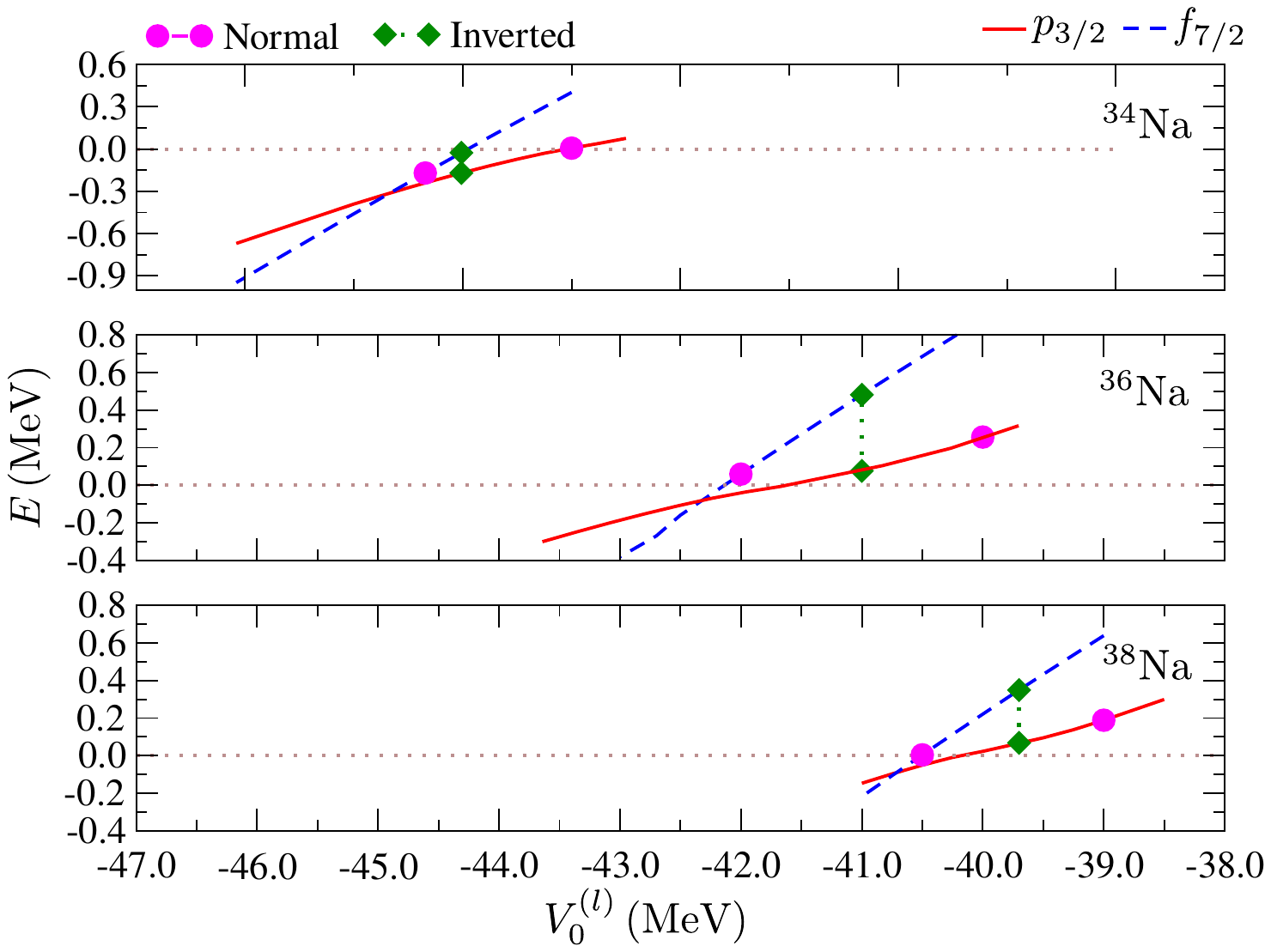}
\caption{The valence single-particle energies of the  neutron $f_{7/2}$ and $p_{3/2}$ orbitals as a function of the central Woods-Saxon depth ($V^{(l)}_0$) for the isotopes $\nucl{34,36,38}{Na}$. Negative energies correspond to bound states, while positive energies indicate unbound continuum (resonance) states. Resonance energies are extracted from a phase-shift analysis.} 
%Solid-filled dots show the parameter choices for the central value of $s_{1n}$ in all three panels. Open points in the upper panel show the choices for the experimental limits on $s_{1n}$, from -$0.67$ to -$0.01\,\text{MeV}$. For simplicity, we assume that the single particle energy of the $p_{3/2}$ orbital in the Normal set is fixed to an energy that is just unbound.}
\label{Fig_2bpot}
\end{figure}

For the even-mass sodium isotopes $\nucl{34,36,38}{Na}$, there are no experimental constraints on the underlying single-particle structure.  The only available information comes from the measurement of the one-neutron separation energy ($s_{1n}$) of $\nucl{34}{Na}$ and the evaluated $s_{1n}$ values of $\nucl{36}{Na}$ and $\nucl{38}{Na}$. The mass measurement shows that $\nucl{34}{Na}$ is weakly bound, with $s_{1n} = 0.17\pm0.50\,\mathrm{MeV}$ \cite{Gaudfroy2012}. The evaluated value has been assigned a larger uncertainty, $s_{1n} = 0.17\pm0.75\,\mathrm{MeV}$ \cite{Wang21}. According to the evaluation \cite{Wang21}, $\nucl{36}{Na}$ and $\nucl{38}{Na}$ are expected to be unbound with $s_{1n} = -0.00\pm0.15\,\mathrm{MeV}$ and $s_{1n} = -0.70\pm0.20\,\mathrm{MeV}$, respectively. In our model, this would just be reflected in the single-particle energies, which we analyse in Fig.~\ref{Fig_2bpot} as a function of the depth of the central Woods-Saxon potential across the three isotopes of interest.

% As stated in the main text, we use range $r_0=1.25\,\mathrm{fm}$ and diffuseness $a=0.75\,\mathrm{fm}$, with a spin–orbit strength $V_{ls}=28.00\,\mathrm{MeV\,fm^2}$.
%\sout{ and thus carry a small uncertainty}
%%%%%%%%%%%%%%%%%%%%%%%%%%%%%%%%%%
We concentrate on two parameter sets: A \textit{normal} set corresponding to the standard shell-model ordering for stable nuclei, where the $f_{7/2}$ orbital lies lower in energy than the $p_{3/2}$ orbital, and an \textit{inverted} one where this order is reversed.
We represent these choices in Fig.~\ref{Fig_2bpot}, where we show the position of these two orbitals (either their binding or resonance energy) as a function of the $l$-dependent central potential strength. 
We see that the energy of the $f_{7/2}$ levels varies strongly with the central potential depth, whereas the $p_{3/2}$ levels are not so sensitive. Therefore, to obtain the normal ordering we need to choose different values of $V_0^{(l)}$ for the $p$ and $f$ states. For $\nucl{34}{Na}$, the central potential depth is chosen so that the $f_{7/2}$ single-particle energy reproduces the central experimental value of $s_{1n}$, while the $p_{3/2}$ state is just unbound and lies above the $f_{7/2}$ level. For $\nucl{36}{Na}$ and $\nucl{38}{Na}$, the central depth is chosen such that the $f_{7/2}$ orbit becomes marginally unbound but still lies lower in energy than the $p_{3/2}$ orbit. 
For  $\nucl{38}{Na}$, this is slightly different from the evaluated $s_{1n}$, but a small upward change does not have a big effect on the results.
We show these normal set solutions in Fig.~\ref{Fig_2bpot} by two magenta dots. %connected by a dash-dotted line. 
Since no experimental constraints exist for the relevant resonances, we set the $f_{7/2}$-$p_{3/2}$ spacing to approximately $0.2\,\mathrm{MeV}$. %(specifically $0.197\,\mathrm{MeV}$ for $\nucl{36}{Na}$ and $0.184\,\mathrm{MeV}$ for $\nucl{38}{Na}$). 
%We could have used a larger spacing, which would enhance the dominance of the lowest orbital in the three-body ground state, but we expect the gap to be rather small. 

These results suggest a more natural choice for the inverted scenario: 
 use a common central depth for $p$ and $f$, that is chosen to support a $p_{3/2}$ ground state across all the three nuclei. This inverted set is represented by green diamonds connected by a dotted line.
Once again, for $\nucl{34}{Na}$, the central potential depth is chosen so that the lowest ($p_{3/2}$) single-particle energy reproduces the central $s_{1n}$ value. This choice simultaneously produces a weakly bound excited $f_{7/2}$ state, consistent with the assumed inversion. For $\nucl{36}{Na}$ and $\nucl{38}{Na}$, the central depth is chosen such that the $p_{3/2}$ orbital is slightly unbound but remains lower in energy than the $f_{7/2}$ orbital. The corresponding $f_{7/2}$-$p_{3/2}$ gaps are then to $0.405\,\mathrm{MeV}$ and $0.281\,\mathrm{MeV}$, for $\nucl{36}{Na}$ and $\nucl{38}{Na}$, respectively. 

Because we adopt relatively small energy gaps between the  $f_{7/2}$ and $p_{3/2}$ orbitals in both the normal and inverted sets, some similarity in the calculated ground-state properties is expected. A significantly larger gap, by contrast, would yield clearly separated theoretical predictions, as seen previously in neutron-rich fluorine isotopes \cite{JSingh2020,GSingh2022}, but would need somewhat unnatural parameters. Our analysis suggest there may be some value to assume an inversion in which the $p_{3/2}$  becomes energetically favoured, also since such an inversion is a hallmark of the shell evolution in similar weakly bound systems. Moreover, the lower angular momentum of the $p_{3/2}$ orbital promotes spatially extended neutron configurations, thereby supporting the emergence of halo structures. Even though there is insufficient experimental data to make such a prediction rigorous, there seems to be a slight preference for this scenario.

%%%%%%%%%%%%%%%%%%%%%%%%%%%%%%%%%%%%%%%%%%%%%%%%%%%%%%%%%%%%%%%%%%%%%%%%%%%%%%%%%%%%%%%%%%%%%%%%%%%%%%%%%%%%%%%%%%%%%%%%%%%%%%%%%%%%%%%%%%%%%%%%%%%%%%%%%%%%%%%%%%%%%%%%%%%%%%%%%%%%%%%%%%%%%%%%%%%%%%%%%%%%%%%%%
\section{Ground state Calculations}\label{GSC}
Using the $\text{core}+n$ potentials defined in the previous section, we compute ground-states and their properties, matter radii and configuration mixing, for the nuclei $\nucl{34,37,39}{Na}$. The nucleus $\nucl{34}{Na}$ is treated as a potential one-neutron halo system, and thus as a $\nucl{33}{Na}+n$ structure, while $\nucl{37}{Na}$ and $\nucl{39}{Na}$ are described as Borromean halo systems with a $\nucl{35}{Na}$ and $\nucl{37}{Na}$ core, respectively. In all three cases the $\nucl{33,35,37}{Na}$ cores are assumed to be inert and are treated as if they have zero angular momentum. A more realistic treatment would require detailed spectroscopic information for $\nucl{34,37,39}{Na}$ and theoretical models of the $\text{core}+n$ interactions that include core excitation and deformation effects.

\begin{figure}[t]
\centering
\includegraphics[width=1.05\linewidth]{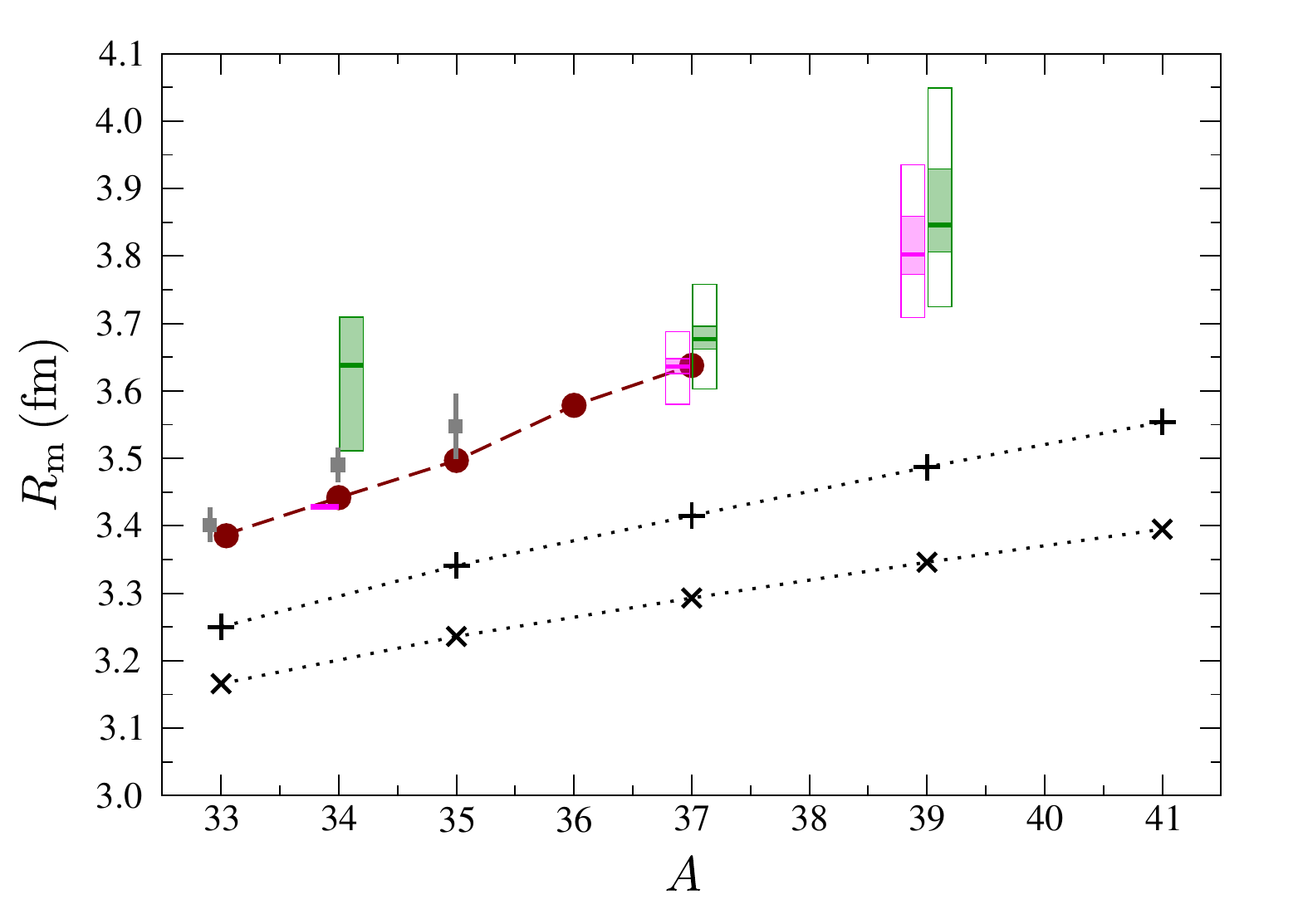}
\caption{Matter radii ($R_m$) of Na isotopes as a function of mass number $A$. The relativistic mean-field (RMF-BCS) (circles) and shell-model results (pluses and crosses), are taken from Refs.~\cite{GENG2004} and \cite{Otsuka2022}, respectively. The gray points (with error bars) are extracted from the preliminary data analysis in Ref.~\cite{Suzuki2014}.
Our results are represented as a band, corresponding to a range of parameters used. Thus, the magenta and green bands
correspond to the normal and inverted parameter sets as shown in Fig.~\ref{Fig_2bpot}, respectively, for different values of $s_{1n}$ ($A=34$) and $s_{2n}$ ($A=37,39$). 
The solid line inside each band is the result obtained with the central experimental or evaluated value of $s_{1n/2n}$. The coloured bands correspond to the full range between the lower and upper limits of the neutron removal energy; the open bands for $A=37,39$ represent a wider range from the deeply bound state with no three-body force (lower limit) to an almost unbound nucleus (upper limit).}
\label{FigmatR}
\end{figure}
%%%%%%%%%%%%%%%%%%%%%%%%%%%%%%%%%%
Since we wish to study a range of values for $s_{1n}$ in  $\nucl{34}{Na}$, we must use a range of $\text{core}+n$ potentials. In the normal set, we only vary the strength in the $l=3$ (i.e., $f_{7/2}$) channel to achieve a shallower and deeper choice, $s_{1n}=0.01$ and $0.67\,\mathrm{MeV}$, respectively. For the inverted set, maintaining a single potential depth $V^{(l)}_0$ for both orbitals is not physically possible at deeper separation energies due to the crossover of the $f_{7/2}$ and $p_{3/2}$ single-particle energy curves. Consequently, while a common $V^{(l)}_0$ suffices for shallow binding ($s_{1n}=0.01\,\mathrm{MeV}$), moving toward deeper binding  ($s_{1n}=0.67\,\mathrm{MeV}$) requires decoupled potential strengths to maintain the inversion. These values of $V^{(l)}_0$ can be easily inferred from the upper panel of Fig.~\ref{Fig_2bpot}. For the case of  $\nucl{36,38}{Na}$, we only use a single set of $\text{core}+n$ potentials, described in Sec.\ref{2bpots} for various choices of $s_{2n}$.

For our three-body model of $\nucl{37}{Na}$ and $\nucl{39}{Na}$ we also need to include an explicit neutron-neutron ($nn$) interaction. We adopt the Gogny-Pires-de Tourreil (GPT) potential \cite{Gogny70} for this purpose, which has been proven effective in three-body calculations for light to medium-mass nuclear systems, ranging from $\nucl{6}{He}$ \cite{JCasal13} to $\nucl{29}{F}$ \cite{JCasal20}. As noted in Ref.~\cite{ZHUKOV1993}, the ground-state properties of $\text{core}+2n$ systems are largely insensitive to the specific choice of the $nn$ interaction, as long as the low-energy $nn$ scattering data are reproduced with sufficient accuracy. The final ingredient of our three-body model is a phenomenological three-body force, $V_{3b}$, introduced to reproduce the ground-state energy ($s_{2n}$), see, e.g., Refs.~\cite{JCasal13,JSingh2020}. In the case of the potential Borromean nuclei $\nucl{37}{Na}$ and $\nucl{39}{Na}$ we need repulsive values of $V_{3b}$ to reproduce the desired $s_{2n}$, see below.

We have truncated the basis as discussed in Sec.~\ref{sec:MF}, choosing $K_{\text{max}}=46$ and $N=20$, which provide converged results for the ground-state properties of $\nucl{37}{Na}$ and $\nucl{39}{Na}$.
The calculation of the matter radii in a few-body model requires as part of the inputs the matter radii of the corresponding cores; in our case, $\nucl{33,35,37}{Na}$.
We adopt core radii $R_\text{core}=3.385$, $3.497$, and $3.638\,\mathrm{fm}$ for $\nucl{33,35,37}{Na}$, respectively, taken from relativistic mean field (RMF-BCS) calculations of Ref.~\cite{GENG2004}. 
Since these values are subject to model uncertainties, our discussion focuses primarily on the relative difference between potential $\text{core}+n$ sets and on the difference in size between the core and the effective two- and three-body systems.

The calculated matter radii for the ground states of $\nucl{34,37,39}{Na}$ obtained with the different sets of $\text{core}+n$ potentials are shown in Fig.~\ref{FigmatR}. As noted earlier in the introduction and Sec.\ref{2bpots}, the experimental and evaluated values for one- and two-neutron separation energies, $s_{1n/2n}$,  carry substantial uncertainties for Na isotopes with $A>32$ \cite{Gaudfroy2012, Wang21}.
For $\nucl{34}{Na}$, we take $s_{1n}=0.17^{+0.50}_{-0.16}\,\mathrm{MeV}$, where the central and upper limit are from experiment \cite{Gaudfroy2012}, and the lower limit reflects the existence of this bound nucleus. The bands shown in Fig.~\ref{FigmatR} represent this full range, with internal line giving the central value.

For $\nucl{37}{Na}$, we use $s_{2n}=0.84\pm0.15\,\mathrm{MeV}$ \cite{Wang21}, which is again represented as a coloured band--much narrower due to the small error quoted. We also analyse much more deeply bound states, which are obtained by reducing $V_{3b}$ to zero (this would overbind the nucleus)--this gives a much lower $R_m$. We also take the upper limit of $R_m$ where the nucleus is bound by only $0.40\,\mathrm{MeV}$. These results are represented as the open boxes in the figure.

Finally, the case of $\nucl{39}{Na}$ is more unusual: the evaluated mass data predict it to be unbound \cite{Wang21}, but recent experimental data unambiguously confirm that it is a bound nucleus, although without an estimate of binding energy \cite{Ahn2022}. We therefore adopt a range of two-neutron separation energies from our recent work \cite{JSingh2024}, $s_{2n}=0.50\pm0.25\,\mathrm{MeV}$. This range is represented in Fig.~\ref{FigmatR} as a coloured band, which is wider than that for $\nucl{37}{Na}$. Similarly, we explore more deeply bound configurations obtained by setting the three-body interaction $V_{3b}$ to zero, which leads to an overbinding of the nucleus and consequently to a substantially smaller  $R_m$. We also consider the upper limit of $R_m$, corresponding to a weakly bound system with $s_{2n}=0.10\,\mathrm{MeV}$. These limiting cases are indicated again by open boxes in the figure.

As shown in Fig.~\ref{FigmatR}, for all systems $\nucl{34,37,39}{Na}$, the standard shell-model ordering for the single particles consistently yields smaller matter radii than those where the single particle states are inverted for similar choices of $s_{1n/2n}$. As expected, the shallowest $s_{1n/2n}$ naturally produces the largest matter radii for all systems, whereas more deeply-bound ground states lead to smaller radii. 
For $\nucl{34}{\text{Na}}$, the normal set contains only a single  $f$-wave component, with a wave function that in not very sensitive to changes in the depth of the Wood-Saxon potential used, and thus shows an almost negligible dependence of the matter radius on $s_{1n}$, as can be seen in  the very thin magenta band. In contrast, the inverted set, with a pure $p$-wave component, exhibits a stronger $s_{1n}$ dependence, resulting in the taller light-green band. This sensitivity reflects the reduced centrifugal barrier for lower angular momentum, which enhances halo formation and increases the matter radius. 
As stated before in Sec.~\ref{2bpots}, for $\nucl{37,39}{Na}$  the energy separation between the $f_{7/2}$ and $p_{3/2}$ orbitals in the normal configuration is relatively small, about $0.2\,\mathrm{MeV}$. In the inverted set, this separation increases slightly to approximately $0.3$-$0.4\,\mathrm{MeV}$ for $\nucl{37,39}{Na}$. This modest splitting is the main reason why the matter radii obtained with the normal and inverted configurations are rather similar. %Tuning the parameters to a larger $f_{7/2}$-$p_{3/2}$ gap would produce a clearer distinction between the two configuration sets over the entire range of $s_{2n}$, as previously demonstrated for the fluorine isotopes \cite{JSingh2020, GSingh2022}, but this would require Wood-Saxon potentials that are more strongly $l$-dependent.
It is worth noting that for $\nucl{37,39}{Na}$, if the $p_{3/2}$ orbit is taken to be fully occupied in the core (as depicted in Fig.~\ref{Fig1})  and therefore excluded from the valence space, our calculations remain close to the normal set. In particular, the matter radii obtained for the smallest $s_{2n}$ values still fall within the white band of the normal set shown in Fig.~\ref{FigmatR}.

\begin{figure}[t]
\centering
\includegraphics[width=0.95\linewidth]{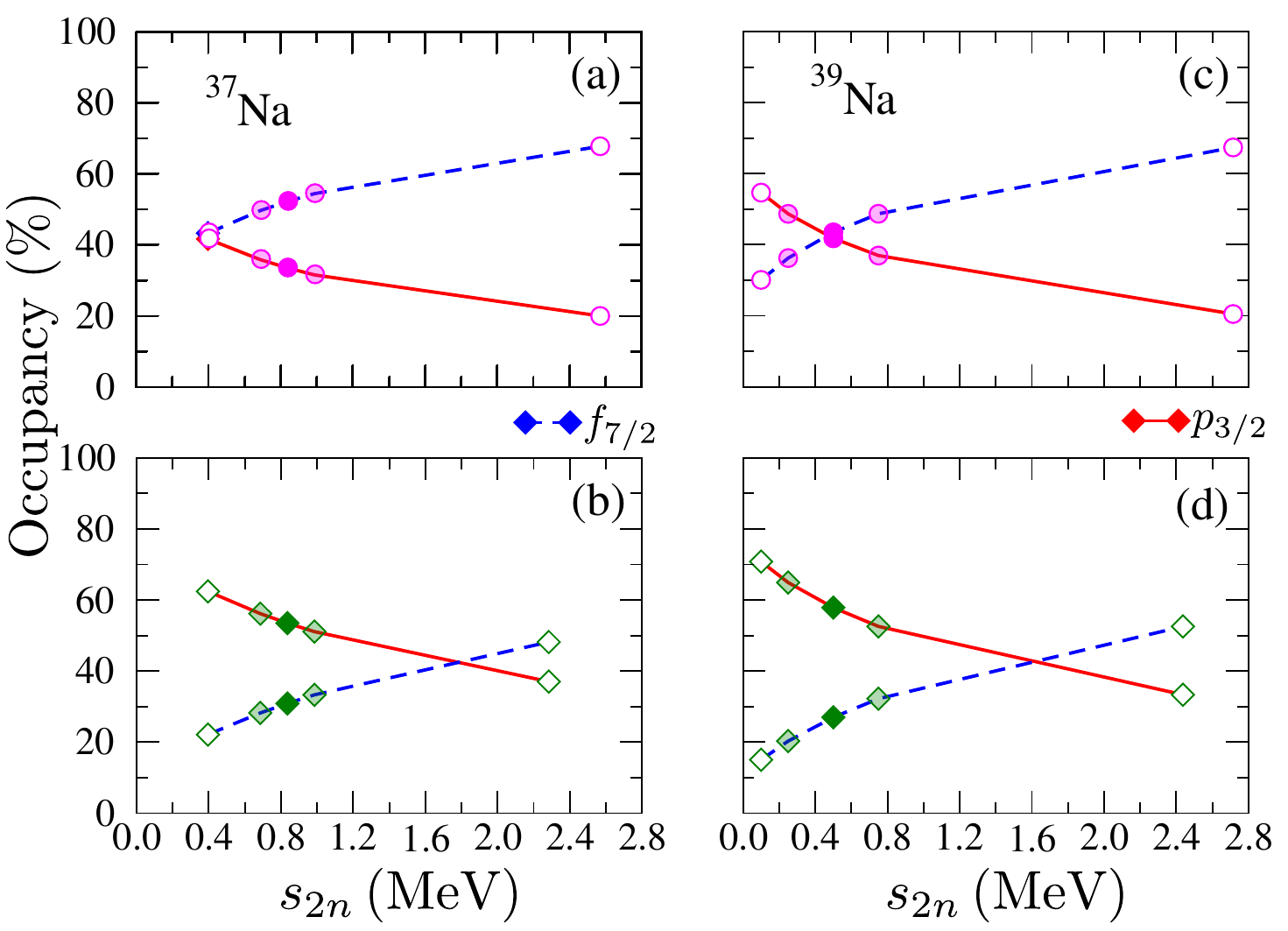}
\caption{Filling fractions of the $(f_{7/2})^2$ (dashed blue) and $(p_{3/2})^2$ (solid red) orbits in $\nucl{37}{Na}$ and $\nucl{39}{Na}$ as a function of two-neutron separation energy $s_{2n}$. Panels (a) and (b) present the results for $\nucl{37}{Na}$, while panels (c) and (d) correspond to 
$\nucl{39}{Na}$. The upper row (a, c) displays calculations using the normal parameter set, and the lower row (b, d) shows those obtained with the inverted set. Solid symbols show the central value,
with transparently filled symbols show the normal limits. Open symbols show the extreme limits (white boxes in Fig.~\ref{FigmatR}). }
\label{Figoccup}
\end{figure}
%%%%%%%%%%%%%%%%%%%%%%%%%%%%%%%%%%
Along with our results, in Fig.~\ref{FigmatR} we also compare to relativistic mean field (RMF-BCS) \cite{GENG2004} and two sets of shell-model calculations \cite{Otsuka2022}.  Both shell-model results include deformation effects, while the results shown as pluses also include neutron excitations into the $pf$ shell (see Ref.~\cite{Otsuka2022} for details). 
Since the core radii in our few-body models are taken from the RMF-BCS framework, our predicted radii are naturally close to those results. It should be noted that the change of trend below $A=35$ and above is something that is reflected most strongly in the inverted scenarios. The nucleus $\nucl{39}{Na}$ is (barely) unbound in the RMF-BCS, but our calculations shown that we can continue the trend of such calculations and predict a bound Borromean nucleus.
The shell-model radii have a smooth dependence on $A$, and clearly do not agree with the inverted scenario we sketch here. However, the choice of single particle energies in such shell model calculations is a difficult problem to solve. Therefore, there is no guarantee that such calculations retain their validity all the way from strongly-bound nuclei to extremely weakly bound ones, especially if they have a halo structure. Figure~\ref{FigmatR} shows also the experimental results from the preliminary analysis of Ref.~\cite{Suzuki2014} for $A=33$-$35$. It is worth noting that these values are somewhat consistent with the RMF-BCS calculation, and so justify our choice of core radii for the computation of matter radii within the present few-body approach.

An important observation is that the sensitivity of the radius to the value of $s_{1n/2n}$ is most pronounced when shell inversion occurs. Nevertheless, owing to the relatively heavy cores, the total radii remain broadly similar across the scenarios considered. This implies that distinguishing between structural scenarios on the basis of radius alone may be experimentally challenging. Additional information from knockout or transfer reactions, capable of constraining the partial-wave composition of the ground state, would be very useful for refining theoretical models and discriminating between the wave functions scenarios discussed in this work.

The nucleus $\nucl{34}{Na}$ is fundamentally different from the other two cases, as it is only a two-body bound system.
As a result, its ground-state wavefunction is a pure single-particle state: it is entirely  $f$-wave for the normal potential set and entirely $p$-wave for the inverted set. A proper treatment of configuration mixing would require a full shell-model calculation or another few- or many-body approach. For instance, core excitations could result in a ground state mixing various single-particle components, similarly to the situation for the halo nucleus $^{11}$Be~\cite{JALay10,Amoro12}.

The wave functions for the two other nuclei, $\nucl{37}{Na}$ and $\nucl{39}{Na}$, is dominated by the occupancies of the $(f_{7/2})^2$ and $(p_{3/2})^2$--other orbits do play a small role. These contributions are shown as functions of $s_{2n}$ as dashed blue and solid red lines, respectively, in Fig.~\ref{Figoccup}, for both sets of $\text{core}+n$ potentials. As expected, shallower (more weakly bound) ground states exhibit a larger $p$-wave occupancy, which in turn results in an increased matter radius, while more deeply bound states yield smaller matter radii for all potential sets. As shown in the lower panel of Fig.~\ref{Figoccup}, reducing the binding energy leads to an increase in the $p$-wave component and a corresponding decrease in the $f$-wave component for each inverted scenario.

For the normal parameter set, shown in the upper panels of Fig.~\ref{Figoccup}, the $f$-wave content is dominant over most of the range of $s_{2n}$. Around $s_{2n}=0.40\,\mathrm{MeV}$ a crossover takes place between the $p$- and $f$-wave occupancy. This behaviour originates from the small energy spacing between the $p$- and $f$-wave orbitals. Interestingly, the percentage of the $p$-wave occupancy never gets smaller than about $20\%$, and thus we can expect some halo-like effects in these odd-$A$ isotopes independent of the scenario.

Clearly, this raises the possibility of finding the appropriate model by looking in detail at the ground-state spectroscopic information, which would constrain the parameter choices allowed. 

\section{$B(E1)$ response}
After discussing the ground-state properties, we now turn to the low-lying electric dipole (E1) response of the Na isotopes. An enhanced low-energy 
$B(E1)$ response is a well-known signature of halo nuclei and is typically extracted from Coulomb dissociation measurements. This is a robust measure: the matter radii discussed in Sec.~\ref{GSC} are more indirectly inferred from high-energy reaction cross-section measurements. 

The dipole strength for transitions from the ground state to continuum states with angular momentum $j$ is given by \begin{equation}B(E1) = |\langle \textit{gs} ||\widehat{O}_{E1}||\varepsilon_n,j\rangle|^2\,,\end{equation}
where in our effective few-body model the electric dipole operator is expressed as
\begin{equation}
    \widehat{O}_{E1} = Ze\sqrt{\frac{2}{A(A+2)}}y Y_{1 M}(\hat{y})\,.
\end{equation}
Since in our calculations the continuum is approximated by discrete pseudo-states each of which contributes a delta function peak with its own dipole strength, these strengths are smoothed by folding with a Poisson kernel, which preserves the integrated total strength~\cite{JCasal13}.

\begin{figure}[t]
\centering
\includegraphics[width=0.95\linewidth]{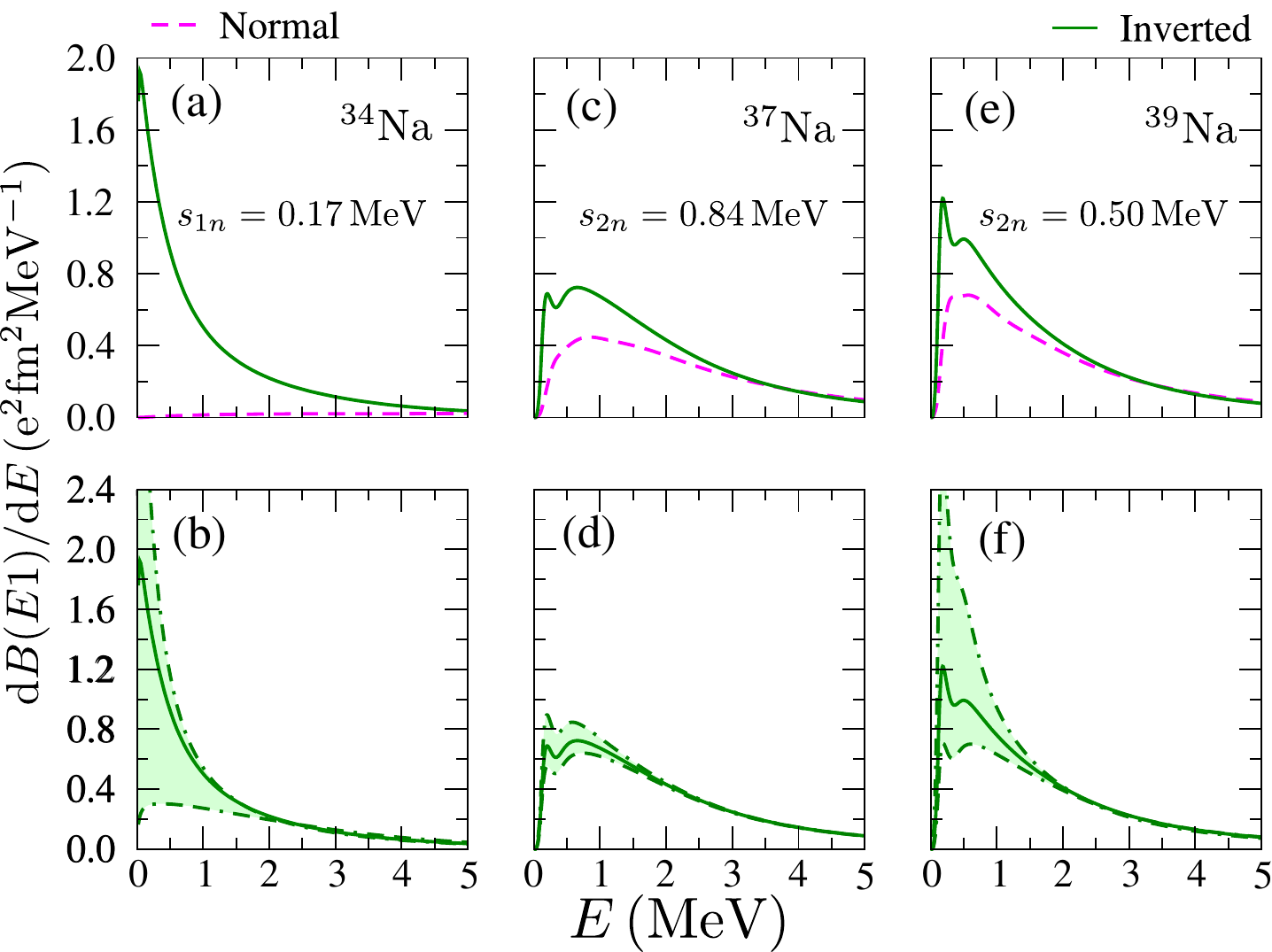}
\caption{$B(E1)$ distributions for $\nucl{34}{Na}$ (panels a, b), $\nucl{37}{Na}$ (panels c, d), and $\nucl{39}{Na}$ (panels e, f). The upper panels (a, c, e) show results calculated at the central experimental/evaluated separation energy for both the normal (dashed magenta line) and inverted (solid green line) sets. The lower panels (b, d, f) display results for the inverted set. The shaded band corresponds to the solid light green band in Fig.~\ref{FigmatR}, for the full range of two-neutron separation energies, and the solid line is for the central value. For details, see main text.}
\label{FigBE1}
\end{figure}
%%%%%%%%%%%%%%%%%%%%%%%%%%%%%%%%%%
%The $1^-$ continuum states are generated using a THO basis with parameters $b=0.7$ fm and $\gamma=1.0$ fm$^{1/2}$, which modify the Gaussian asymptotic behavior of the harmonic-oscillator functions into an exponential fall-off. The ratio $\gamma/b$ governs the hyperradial extension of the basis. 
There is no simple convergence criterion as for the ground state. In this case, we use stability of the $B(E1)$ distribution, which is less straightforward to judge. We find that converged results are achieved with $N=24$ and $K_{\rm max}=38$.
For simplicity, we have assumed that the three-body force used for the ground state also applies here--in principle, this could be state dependent, but we lack the experimental data to make such a judgement. 

Our results for the $B(E1)$ distributions are displayed in Fig.~\ref{FigBE1} for $\nucl{34}{Na}$ (panels a and b), $\nucl{37}{Na}$ (panels c and d), and $\nucl{39}{Na}$ (panels e and f).
In the upper panel of Fig.~\ref{FigBE1}, we show the calculated $B(E1)$ distributions for $\nucl{34,37,39}{Na}$ using the central experimental and evaluated separation energies $s_{1n}=0.17\,\mathrm{MeV}$ for $\nucl{34}{Na}$, $s_{2n}=0.84\,\mathrm{MeV}$ for $\nucl{37}{Na}$, and $s_{2n}=0.50\,\mathrm{MeV}$ for $\nucl{39}{Na}$. The calculations are performed for $\text{core}+n$ potential for  both normal and inverted single-particle levels (central value for the removal energies only). The inverted ordering gives rise to a pronounced low-energy peak in the dipole strength distribution in all three nuclei, showing again that this corresponds to a halo structure. For the same separation energies, the normal set consistently produces (much) smaller $B(E1)$ strength. This difference is most significant for the one-neutron halo nucleus $\nucl{34}{Na}$, reflecting the single-particle character of its ground state within our simple model. In the Borromean halos, the visible enhancement arises due to the larger $p$-wave component generated by the inverted potential in the ground-state, which increases the matter radius and, consequently, the dipole strength.

In the lower panel of Fig.~\ref{FigBE1}, we present the $B(E1)$ distributions for $\nucl{34,37,39}{Na}$ calculated with different separation energies. As seen in Fig.~\ref{FigBE1}, the shallowest ground states, characterised by the largest $p$-wave occupancy and matter radius, exhibit the strongest dipole response. The light green shaded bands show the full range of separation energies between the lower (dash-dotted line), and upper (dotted double–dashed line) separation-energy limits. The solid green line marks the central value, and for it the integrated dipole strength up to $5\,\mathrm{MeV}$ amounts to a substantial fraction ($~85\%$) of the corresponding cluster sum-rule limits. These results provide clear evidence for a pronounced halo structure in the shallowest configurations of all three $\nucl{34,37,39}{Na}$ isotopes. The wiggles in the $B(E1)$ distributions may be numerical artifacts of the smoothing procedure. This suggests a more rigorous identification of three-body resonances is required; this will be addressed in future work.

Thus,  experiments that study the $B(E1)$ strength in the sodium nuclei should be able to distinguish between an inverted and normal single particle ordering. This would allow unambiguous evidence of possible halo formation.
\section{Conclusions and Outlook}\label{sec:conc}

In this work, we have investigated the crucial role of the position of the $f_{7/2}$ and $p_{3/2}$ single-particle orbitals for the ground-state of neutron-rich sodium isotopes.
From the available systematics, it seems likely that the order inverts as we increase the number of neutrons. 
Since there is limited experimental information for Na isotopes with mass number $A>33$ to constrain the relevant parameters, we have performed few-body calculations for a number of parameter sets, both with and without such an inversion. We use this to obtain the ground state matter radius, configuration mixing, and electric dipole response for the halo candidates $\nucl{34,37,39}{Na}$. As expected, the observables are sensitive to the underlying shell evolution driven by the $f_{7/2}$-$p_{3/2}$ inversion, and thus an experimental study of these observables would provide invaluable information to confirm whether they sustain halo formation.

Our results clearly show that neutron halo formation is strongly favoured in the inverted scenarios--and from systematics, we expect that a softening gap and potentially an inversion will play a role in these nuclei. The enhancement of the matter radii and the emergence of pronounced low-energy dipole strength will provide consistent signatures of spatially-extended weakly-bound neutrons. This does not preclude halos in the normal scenario, even though radii and response strengths are found to be smaller. Despite its simplicity, our few-body approach captures the core physics of halo nuclei, including two-neutron correlations that drive the weak binding of neutrons in loosely bound orbits in Borromean systems. %This constitutes a clear advantage over mean-field approaches, where such correlations are absent or only effectively included.}

The present study shows that sodium isotopes are an interesting area to study shell evolution and halo phenomena. Future experimental measurements of interaction cross sections, breakup reactions, and dipole response for Na isotopes with  $A>33$ will be crucial for validating the suggested potential orbital inversion and its consequences for halo structure. On the theoretical side, extending the present few-body framework to include core deformation, excitations, and full details of angular momentum structure will further improve the quantitative understanding of these exotic systems.
%%%%%%%%%%%%%%%%%%%%%%%%%%%%%%%%%%%%%%%%%%%%%%%%%%%%%%%%%%%%%%%%%%%%%%%%%%%%%%%%%%%%%%%%%%%%%%%%%%%%%%%%%%%%%%%%%%%%%%%%%%%%%%%%%%%%%%%%%%%%%%%%%%%%%%%%%%
\section*{Acknowledgments}
This work is dedicated to the memory of \textit{Prof.~Andrea Vitturi}. His wisdom and guidance profoundly shaped the scientific journeys of LF, JC, and JS, and his inspiration continues to guide us. 
We thank Juan Lois-Fuentes for pointing us towards studying  $\nucl{37}{Na}$, Olivier Sorlin and Takaharu Otsuka for helpful discussions, and Naohiko Otsuka for clarifying the status of the experimental analysis and for providing the digitised experimental data shown in Fig.~\ref{FigmatR}. We thank \textit{INFN Padova} for supporting a research visit by JS and JC in May 2025, during which this project was initiated. 
This work was supported by the UK Science and Technology Funding Council [grant number ST/Y000323/1] (JS and NRW); by the Spanish MICIU/AEI/10.13039/501100011033 and by ERDF/EU [project No.~PID2023-146401NB-I00] (JC); and by INFN-Monstre (LF).
%%%%%%%%%%%%%%%%%%%%%%%%%%%%%%%%%%%%%%%%%%%%%%%%%%%%%%%%%%%%%%%%%%%%%%%%%%%%%%%%%%%%%%%%%%%%%%%%%%%%%%%%%%%%%%%%%%%%%%%%%%%%%%%%%%%%%%%%%%%%%%%%%%%%%%%%%%
\bibliography{ref}	

%merlin.mbs apsrev4-1.bst 2010-07-25 4.21a (PWD, AO, DPC) hacked
%Control: key (0)
%Control: author (72) initials jnrlst
%Control: editor formatted (1) identically to author
%Control: production of article title (-1) disabled
%Control: page (0) single
%Control: year (1) truncated
%Control: production of eprint (0) enabled
\begin{thebibliography}{32}%
\makeatletter
\providecommand \@ifxundefined [1]{%
 \@ifx{#1\undefined}
}%
\providecommand \@ifnum [1]{%
 \ifnum #1\expandafter \@firstoftwo
 \else \expandafter \@secondoftwo
 \fi
}%
\providecommand \@ifx [1]{%
 \ifx #1\expandafter \@firstoftwo
 \else \expandafter \@secondoftwo
 \fi
}%
\providecommand \natexlab [1]{#1}%
\providecommand \enquote  [1]{``#1''}%
\providecommand \bibnamefont  [1]{#1}%
\providecommand \bibfnamefont [1]{#1}%
\providecommand \citenamefont [1]{#1}%
\providecommand \href@noop [0]{\@secondoftwo}%
\providecommand \href [0]{\begingroup \@sanitize@url \@href}%
\providecommand \@href[1]{\@@startlink{#1}\@@href}%
\providecommand \@@href[1]{\endgroup#1\@@endlink}%
\providecommand \@sanitize@url [0]{\catcode `\\12\catcode `\$12\catcode `\&12\catcode `\#12\catcode `\^12\catcode `\_12\catcode `\%12\relax}%
\providecommand \@@startlink[1]{}%
\providecommand \@@endlink[0]{}%
\providecommand \url  [0]{\begingroup\@sanitize@url \@url }%
\providecommand \@url [1]{\endgroup\@href {#1}{\urlprefix }}%
\providecommand \urlprefix  [0]{URL }%
\providecommand \Eprint [0]{\href }%
\providecommand \doibase [0]{http://dx.doi.org/}%
\providecommand \selectlanguage [0]{\@gobble}%
\providecommand \bibinfo  [0]{\@secondoftwo}%
\providecommand \bibfield  [0]{\@secondoftwo}%
\providecommand \translation [1]{[#1]}%
\providecommand \BibitemOpen [0]{}%
\providecommand \bibitemStop [0]{}%
\providecommand \bibitemNoStop [0]{.\EOS\space}%
\providecommand \EOS [0]{\spacefactor3000\relax}%
\providecommand \BibitemShut  [1]{\csname bibitem#1\endcsname}%
\let\auto@bib@innerbib\@empty
%</preamble>
\bibitem [{\citenamefont {Nunes}(2021)}]{Filomena2021}%
  \BibitemOpen
  \bibfield  {author} {\bibinfo {author} {\bibfnamefont {F.~M.}\ \bibnamefont {Nunes}},\ }\href {\doibase 10.1063/PT.3.4748} {\bibfield  {journal} {\bibinfo  {journal} {Physics Today}\ }\textbf {\bibinfo {volume} {74}},\ \bibinfo {pages} {34} (\bibinfo {year} {2021})}\BibitemShut {NoStop}%
\bibitem [{\citenamefont {Crawford}\ \emph {et~al.}(2024)\citenamefont {Crawford}, \citenamefont {Fossez}, \citenamefont {König},\ and\ \citenamefont {Spyrou}}]{Crawford2024}%
  \BibitemOpen
  \bibfield  {author} {\bibinfo {author} {\bibfnamefont {H.~L.}\ \bibnamefont {Crawford}}, \bibinfo {author} {\bibfnamefont {K.}~\bibnamefont {Fossez}}, \bibinfo {author} {\bibfnamefont {S.}~\bibnamefont {König}}, \ and\ \bibinfo {author} {\bibfnamefont {A.}~\bibnamefont {Spyrou}},\ }\href {\doibase 10.1146/annurev-nucl-121423-091501} {\bibfield  {journal} {\bibinfo  {journal} {Annual Review of Nuclear and Particle Science}\ }\textbf {\bibinfo {volume} {74}},\ \bibinfo {pages} {141} (\bibinfo {year} {2024})}\BibitemShut {NoStop}%
\bibitem [{\citenamefont {Brown}\ \emph {et~al.}(2025)\citenamefont {Brown}, \citenamefont {Gade}, \citenamefont {Stroberg}, \citenamefont {Escher}, \citenamefont {Fossez}, \citenamefont {Giuliani}, \citenamefont {Hoffman}, \citenamefont {Nazarewicz}, \citenamefont {Seng}, \citenamefont {Sorensen}, \citenamefont {Vassh}, \citenamefont {Bazin}, \citenamefont {Brown}, \citenamefont {Caprio}, \citenamefont {Crawford}, \citenamefont {Danielewicz}, \citenamefont {Drischler}, \citenamefont {García~Ruiz} \emph {et~al.}}]{Brown2025_FRIB}%
  \BibitemOpen
  \bibfield  {author} {\bibinfo {author} {\bibfnamefont {B.}~\bibnamefont {Brown}}, \bibinfo {author} {\bibfnamefont {A.}~\bibnamefont {Gade}}, \bibinfo {author} {\bibfnamefont {S.}~\bibnamefont {Stroberg}}, \bibinfo {author} {\bibfnamefont {J.}~\bibnamefont {Escher}}, \bibinfo {author} {\bibfnamefont {K.}~\bibnamefont {Fossez}}, \bibinfo {author} {\bibfnamefont {P.}~\bibnamefont {Giuliani}}, \bibinfo {author} {\bibfnamefont {C.}~\bibnamefont {Hoffman}}, \bibinfo {author} {\bibfnamefont {W.}~\bibnamefont {Nazarewicz}}, \bibinfo {author} {\bibfnamefont {C.}~\bibnamefont {Seng}}, \bibinfo {author} {\bibfnamefont {A.}~\bibnamefont {Sorensen}}, \bibinfo {author} {\bibfnamefont {N.}~\bibnamefont {Vassh}}, \bibinfo {author} {\bibfnamefont {D.}~\bibnamefont {Bazin}}, \bibinfo {author} {\bibfnamefont {W.}~\bibnamefont {Brown}}, \bibinfo {author} {\bibfnamefont {M.}~\bibnamefont {Caprio}}, \bibinfo {author} {\bibfnamefont {H.}~\bibnamefont {Crawford}}, \bibinfo {author} {\bibfnamefont {P.}~\bibnamefont {Danielewicz}},
  \bibinfo {author} {\bibfnamefont {C.}~\bibnamefont {Drischler}}, \bibinfo {author} {\bibfnamefont {R.}~\bibnamefont {García~Ruiz}},  \emph {et~al.},\ }\href {\doibase 10.1088/1361-6471/adb449} {\bibfield  {journal} {\bibinfo  {journal} {Journal of Physics G: Nuclear and Particle Physics}\ }\textbf {\bibinfo {volume} {52}},\ \bibinfo {pages} {050501} (\bibinfo {year} {2025})}\BibitemShut {NoStop}%
\bibitem [{\citenamefont {Hamamoto}(2007)}]{Hammamoto2007}%
  \BibitemOpen
  \bibfield  {author} {\bibinfo {author} {\bibfnamefont {I.}~\bibnamefont {Hamamoto}},\ }\href {\doibase 10.1103/PhysRevC.76.054319} {\bibfield  {journal} {\bibinfo  {journal} {Phys. Rev. C}\ }\textbf {\bibinfo {volume} {76}},\ \bibinfo {pages} {054319} (\bibinfo {year} {2007})}\BibitemShut {NoStop}%
\bibitem [{\citenamefont {Hamamoto}(2012)}]{Hammamoto2012}%
  \BibitemOpen
  \bibfield  {author} {\bibinfo {author} {\bibfnamefont {I.}~\bibnamefont {Hamamoto}},\ }\href {\doibase 10.1103/PhysRevC.85.064329} {\bibfield  {journal} {\bibinfo  {journal} {Phys. Rev. C}\ }\textbf {\bibinfo {volume} {85}},\ \bibinfo {pages} {064329} (\bibinfo {year} {2012})}\BibitemShut {NoStop}%
\bibitem [{\citenamefont {Kobayashi}\ \emph {et~al.}(2016)\citenamefont {Kobayashi}, \citenamefont {Nakamura}, \citenamefont {Kondo}, \citenamefont {Tostevin}, \citenamefont {Aoi}, \citenamefont {Baba}, \citenamefont {Barthelemy}, \citenamefont {Famiano}, \citenamefont {Fukuda}, \citenamefont {Inabe}, \citenamefont {Ishihara}, \citenamefont {Kanungo}, \citenamefont {Kim}, \citenamefont {Kubo}, \citenamefont {Lee}, \citenamefont {Lee}, \citenamefont {Matsushita}, \citenamefont {Motobayashi}, \citenamefont {Ohnishi}, \citenamefont {Orr}, \citenamefont {Otsu}, \citenamefont {Sako}, \citenamefont {Sakurai}, \citenamefont {Satou}, \citenamefont {Sumikama}, \citenamefont {Takeda}, \citenamefont {Takeuchi}, \citenamefont {Tanaka}, \citenamefont {Togano},\ and\ \citenamefont {Yoneda}}]{Kobayashi2016}%
  \BibitemOpen
  \bibfield  {author} {\bibinfo {author} {\bibfnamefont {N.}~\bibnamefont {Kobayashi}}, \bibinfo {author} {\bibfnamefont {T.}~\bibnamefont {Nakamura}}, \bibinfo {author} {\bibfnamefont {Y.}~\bibnamefont {Kondo}}, \bibinfo {author} {\bibfnamefont {J.~A.}\ \bibnamefont {Tostevin}}, \bibinfo {author} {\bibfnamefont {N.}~\bibnamefont {Aoi}}, \bibinfo {author} {\bibfnamefont {H.}~\bibnamefont {Baba}}, \bibinfo {author} {\bibfnamefont {R.}~\bibnamefont {Barthelemy}}, \bibinfo {author} {\bibfnamefont {M.~A.}\ \bibnamefont {Famiano}}, \bibinfo {author} {\bibfnamefont {N.}~\bibnamefont {Fukuda}}, \bibinfo {author} {\bibfnamefont {N.}~\bibnamefont {Inabe}}, \bibinfo {author} {\bibfnamefont {M.}~\bibnamefont {Ishihara}}, \bibinfo {author} {\bibfnamefont {R.}~\bibnamefont {Kanungo}}, \bibinfo {author} {\bibfnamefont {S.}~\bibnamefont {Kim}}, \bibinfo {author} {\bibfnamefont {T.}~\bibnamefont {Kubo}}, \bibinfo {author} {\bibfnamefont {G.~S.}\ \bibnamefont {Lee}}, \bibinfo {author} {\bibfnamefont {H.~S.}\ \bibnamefont
  {Lee}}, \bibinfo {author} {\bibfnamefont {M.}~\bibnamefont {Matsushita}}, \bibinfo {author} {\bibfnamefont {T.}~\bibnamefont {Motobayashi}}, \bibinfo {author} {\bibfnamefont {T.}~\bibnamefont {Ohnishi}}, \bibinfo {author} {\bibfnamefont {N.~A.}\ \bibnamefont {Orr}}, \bibinfo {author} {\bibfnamefont {H.}~\bibnamefont {Otsu}}, \bibinfo {author} {\bibfnamefont {T.}~\bibnamefont {Sako}}, \bibinfo {author} {\bibfnamefont {H.}~\bibnamefont {Sakurai}}, \bibinfo {author} {\bibfnamefont {Y.}~\bibnamefont {Satou}}, \bibinfo {author} {\bibfnamefont {T.}~\bibnamefont {Sumikama}}, \bibinfo {author} {\bibfnamefont {H.}~\bibnamefont {Takeda}}, \bibinfo {author} {\bibfnamefont {S.}~\bibnamefont {Takeuchi}}, \bibinfo {author} {\bibfnamefont {R.}~\bibnamefont {Tanaka}}, \bibinfo {author} {\bibfnamefont {Y.}~\bibnamefont {Togano}}, \ and\ \bibinfo {author} {\bibfnamefont {K.}~\bibnamefont {Yoneda}},\ }\href {\doibase 10.1103/PhysRevC.93.014613} {\bibfield  {journal} {\bibinfo  {journal} {Phys. Rev. C}\ }\textbf {\bibinfo
  {volume} {93}},\ \bibinfo {pages} {014613} (\bibinfo {year} {2016})}\BibitemShut {NoStop}%
\bibitem [{\citenamefont {Kahlbow}\ \emph {et~al.}(2024)\citenamefont {Kahlbow}, \citenamefont {Aumann}, \citenamefont {Sorlin}, \citenamefont {Kondo}, \citenamefont {Nakamura}, \citenamefont {Nowacki}, \citenamefont {Revel}, \citenamefont {Achouri}, \citenamefont {Al~Falou}, \citenamefont {Atar}, \citenamefont {Baba}, \citenamefont {Boretzky}, \citenamefont {Caesar}, \citenamefont {Calvet}, \citenamefont {Chae}, \citenamefont {Chiga}, \citenamefont {Corsi}, \citenamefont {Delaunay}, \citenamefont {Delbart}, \citenamefont {Deshayes}, \citenamefont {Dombr\'adi}, \citenamefont {Douma}, \citenamefont {Elekes}, \citenamefont {Ga\ifmmode \check{s}\else \v{s}\fi{}pari\ifmmode~\acute{c}\else \'{c}\fi{}}, \citenamefont {Gheller}, \citenamefont {Gibelin}, \citenamefont {Gillibert}, \citenamefont {Harakeh}, \citenamefont {Hirayama}, \citenamefont {Holl}, \citenamefont {Horvat}, \citenamefont {Horv\'ath}, \citenamefont {Hwang}, \citenamefont {Isobe}, \citenamefont {Kalantar-Nayestanaki}, \citenamefont {Kawase},
  \citenamefont {Kim}, \citenamefont {Kisamori}, \citenamefont {Kobayashi}, \citenamefont {K\"orper}, \citenamefont {Koyama}, \citenamefont {Kuti}, \citenamefont {Lapoux}, \citenamefont {Lindberg}, \citenamefont {Marqu\'es}, \citenamefont {Masuoka}, \citenamefont {Mayer}, \citenamefont {Miki}, \citenamefont {Murakami}, \citenamefont {Najafi}, \citenamefont {Nakano}, \citenamefont {Nakatsuka}, \citenamefont {Nilsson}, \citenamefont {Obertelli}, \citenamefont {Orr}, \citenamefont {Otsu}, \citenamefont {Ozaki}, \citenamefont {Panin}, \citenamefont {Paschalis}, \citenamefont {Rossi}, \citenamefont {Saito}, \citenamefont {Saito}, \citenamefont {Sasano}, \citenamefont {Sato}, \citenamefont {Satou}, \citenamefont {Scheit}, \citenamefont {Schindler}, \citenamefont {Schrock}, \citenamefont {Shikata}, \citenamefont {Shimada}, \citenamefont {Shimizu}, \citenamefont {Simon}, \citenamefont {Sohler}, \citenamefont {Stuhl}, \citenamefont {Takeuchi}, \citenamefont {Tanaka}, \citenamefont {Thoennessen}, \citenamefont
  {T\"ornqvist}, \citenamefont {Togano}, \citenamefont {Tomai}, \citenamefont {Tscheuschner}, \citenamefont {Tsubota}, \citenamefont {Uesaka}, \citenamefont {Wang}, \citenamefont {Yang}, \citenamefont {Yasuda},\ and\ \citenamefont {Yoneda}}]{Kahlbow2024}%
  \BibitemOpen
  \bibfield  {author} {\bibinfo {author} {\bibfnamefont {J.}~\bibnamefont {Kahlbow}}, \bibinfo {author} {\bibfnamefont {T.}~\bibnamefont {Aumann}}, \bibinfo {author} {\bibfnamefont {O.}~\bibnamefont {Sorlin}}, \bibinfo {author} {\bibfnamefont {Y.}~\bibnamefont {Kondo}}, \bibinfo {author} {\bibfnamefont {T.}~\bibnamefont {Nakamura}}, \bibinfo {author} {\bibfnamefont {F.}~\bibnamefont {Nowacki}}, \bibinfo {author} {\bibfnamefont {A.}~\bibnamefont {Revel}}, \bibinfo {author} {\bibfnamefont {N.~L.}\ \bibnamefont {Achouri}}, \bibinfo {author} {\bibfnamefont {H.}~\bibnamefont {Al~Falou}}, \bibinfo {author} {\bibfnamefont {L.}~\bibnamefont {Atar}}, \bibinfo {author} {\bibfnamefont {H.}~\bibnamefont {Baba}}, \bibinfo {author} {\bibfnamefont {K.}~\bibnamefont {Boretzky}}, \bibinfo {author} {\bibfnamefont {C.}~\bibnamefont {Caesar}}, \bibinfo {author} {\bibfnamefont {D.}~\bibnamefont {Calvet}}, \bibinfo {author} {\bibfnamefont {H.}~\bibnamefont {Chae}}, \bibinfo {author} {\bibfnamefont {N.}~\bibnamefont {Chiga}},
  \bibinfo {author} {\bibfnamefont {A.}~\bibnamefont {Corsi}}, \bibinfo {author} {\bibfnamefont {F.}~\bibnamefont {Delaunay}}, \bibinfo {author} {\bibfnamefont {A.}~\bibnamefont {Delbart}}, \bibinfo {author} {\bibfnamefont {Q.}~\bibnamefont {Deshayes}}, \bibinfo {author} {\bibfnamefont {Z.}~\bibnamefont {Dombr\'adi}}, \bibinfo {author} {\bibfnamefont {C.~A.}\ \bibnamefont {Douma}}, \bibinfo {author} {\bibfnamefont {Z.}~\bibnamefont {Elekes}}, \bibinfo {author} {\bibfnamefont {I.}~\bibnamefont {Ga\ifmmode \check{s}\else \v{s}\fi{}pari\ifmmode~\acute{c}\else \'{c}\fi{}}}, \bibinfo {author} {\bibfnamefont {J.-M.}\ \bibnamefont {Gheller}}, \bibinfo {author} {\bibfnamefont {J.}~\bibnamefont {Gibelin}}, \bibinfo {author} {\bibfnamefont {A.}~\bibnamefont {Gillibert}}, \bibinfo {author} {\bibfnamefont {M.~N.}\ \bibnamefont {Harakeh}}, \bibinfo {author} {\bibfnamefont {A.}~\bibnamefont {Hirayama}}, \bibinfo {author} {\bibfnamefont {M.}~\bibnamefont {Holl}}, \bibinfo {author} {\bibfnamefont {A.}~\bibnamefont {Horvat}},
  \bibinfo {author} {\bibfnamefont {A.}~\bibnamefont {Horv\'ath}}, \bibinfo {author} {\bibfnamefont {J.~W.}\ \bibnamefont {Hwang}}, \bibinfo {author} {\bibfnamefont {T.}~\bibnamefont {Isobe}}, \bibinfo {author} {\bibfnamefont {N.}~\bibnamefont {Kalantar-Nayestanaki}}, \bibinfo {author} {\bibfnamefont {S.}~\bibnamefont {Kawase}}, \bibinfo {author} {\bibfnamefont {S.}~\bibnamefont {Kim}}, \bibinfo {author} {\bibfnamefont {K.}~\bibnamefont {Kisamori}}, \bibinfo {author} {\bibfnamefont {T.}~\bibnamefont {Kobayashi}}, \bibinfo {author} {\bibfnamefont {D.}~\bibnamefont {K\"orper}}, \bibinfo {author} {\bibfnamefont {S.}~\bibnamefont {Koyama}}, \bibinfo {author} {\bibfnamefont {I.}~\bibnamefont {Kuti}}, \bibinfo {author} {\bibfnamefont {V.}~\bibnamefont {Lapoux}}, \bibinfo {author} {\bibfnamefont {S.}~\bibnamefont {Lindberg}}, \bibinfo {author} {\bibfnamefont {F.~M.}\ \bibnamefont {Marqu\'es}}, \bibinfo {author} {\bibfnamefont {S.}~\bibnamefont {Masuoka}}, \bibinfo {author} {\bibfnamefont {J.}~\bibnamefont {Mayer}},
  \bibinfo {author} {\bibfnamefont {K.}~\bibnamefont {Miki}}, \bibinfo {author} {\bibfnamefont {T.}~\bibnamefont {Murakami}}, \bibinfo {author} {\bibfnamefont {M.}~\bibnamefont {Najafi}}, \bibinfo {author} {\bibfnamefont {K.}~\bibnamefont {Nakano}}, \bibinfo {author} {\bibfnamefont {N.}~\bibnamefont {Nakatsuka}}, \bibinfo {author} {\bibfnamefont {T.}~\bibnamefont {Nilsson}}, \bibinfo {author} {\bibfnamefont {A.}~\bibnamefont {Obertelli}}, \bibinfo {author} {\bibfnamefont {N.~A.}\ \bibnamefont {Orr}}, \bibinfo {author} {\bibfnamefont {H.}~\bibnamefont {Otsu}}, \bibinfo {author} {\bibfnamefont {T.}~\bibnamefont {Ozaki}}, \bibinfo {author} {\bibfnamefont {V.}~\bibnamefont {Panin}}, \bibinfo {author} {\bibfnamefont {S.}~\bibnamefont {Paschalis}}, \bibinfo {author} {\bibfnamefont {D.~M.}\ \bibnamefont {Rossi}}, \bibinfo {author} {\bibfnamefont {A.~T.}\ \bibnamefont {Saito}}, \bibinfo {author} {\bibfnamefont {T.}~\bibnamefont {Saito}}, \bibinfo {author} {\bibfnamefont {M.}~\bibnamefont {Sasano}}, \bibinfo {author}
  {\bibfnamefont {H.}~\bibnamefont {Sato}}, \bibinfo {author} {\bibfnamefont {Y.}~\bibnamefont {Satou}}, \bibinfo {author} {\bibfnamefont {H.}~\bibnamefont {Scheit}}, \bibinfo {author} {\bibfnamefont {F.}~\bibnamefont {Schindler}}, \bibinfo {author} {\bibfnamefont {P.}~\bibnamefont {Schrock}}, \bibinfo {author} {\bibfnamefont {M.}~\bibnamefont {Shikata}}, \bibinfo {author} {\bibfnamefont {K.}~\bibnamefont {Shimada}}, \bibinfo {author} {\bibfnamefont {Y.}~\bibnamefont {Shimizu}}, \bibinfo {author} {\bibfnamefont {H.}~\bibnamefont {Simon}}, \bibinfo {author} {\bibfnamefont {D.}~\bibnamefont {Sohler}}, \bibinfo {author} {\bibfnamefont {L.}~\bibnamefont {Stuhl}}, \bibinfo {author} {\bibfnamefont {S.}~\bibnamefont {Takeuchi}}, \bibinfo {author} {\bibfnamefont {M.}~\bibnamefont {Tanaka}}, \bibinfo {author} {\bibfnamefont {M.}~\bibnamefont {Thoennessen}}, \bibinfo {author} {\bibfnamefont {H.}~\bibnamefont {T\"ornqvist}}, \bibinfo {author} {\bibfnamefont {Y.}~\bibnamefont {Togano}}, \bibinfo {author} {\bibfnamefont
  {T.}~\bibnamefont {Tomai}}, \bibinfo {author} {\bibfnamefont {J.}~\bibnamefont {Tscheuschner}}, \bibinfo {author} {\bibfnamefont {J.}~\bibnamefont {Tsubota}}, \bibinfo {author} {\bibfnamefont {T.}~\bibnamefont {Uesaka}}, \bibinfo {author} {\bibfnamefont {H.}~\bibnamefont {Wang}}, \bibinfo {author} {\bibfnamefont {Z.}~\bibnamefont {Yang}}, \bibinfo {author} {\bibfnamefont {M.}~\bibnamefont {Yasuda}}, \ and\ \bibinfo {author} {\bibfnamefont {K.}~\bibnamefont {Yoneda}} (\bibinfo {collaboration} {SAMURAI21-NeuLAND Collaboration}),\ }\href {\doibase 10.1103/PhysRevLett.133.082501} {\bibfield  {journal} {\bibinfo  {journal} {Phys. Rev. Lett.}\ }\textbf {\bibinfo {volume} {133}},\ \bibinfo {pages} {082501} (\bibinfo {year} {2024})}\BibitemShut {NoStop}%
\bibitem [{\citenamefont {Wang}\ \emph {et~al.}(2021)\citenamefont {Wang}, \citenamefont {Huang}, \citenamefont {Kondev}, \citenamefont {Audi},\ and\ \citenamefont {Naimi}}]{Wang21}%
  \BibitemOpen
  \bibfield  {author} {\bibinfo {author} {\bibfnamefont {M.}~\bibnamefont {Wang}}, \bibinfo {author} {\bibfnamefont {W.}~\bibnamefont {Huang}}, \bibinfo {author} {\bibfnamefont {F.}~\bibnamefont {Kondev}}, \bibinfo {author} {\bibfnamefont {G.}~\bibnamefont {Audi}}, \ and\ \bibinfo {author} {\bibfnamefont {S.}~\bibnamefont {Naimi}},\ }\href {\doibase 10.1088/1674-1137/abddaf} {\bibfield  {journal} {\bibinfo  {journal} {Chinese Physics C}\ }\textbf {\bibinfo {volume} {45}},\ \bibinfo {pages} {030003} (\bibinfo {year} {2021})}\BibitemShut {NoStop}%
\bibitem [{\citenamefont {Gade}\ \emph {et~al.}(2011)\citenamefont {Gade}, \citenamefont {Bazin}, \citenamefont {Brown}, \citenamefont {Campbell}, \citenamefont {Cook}, \citenamefont {Ettenauer}, \citenamefont {Glasmacher}, \citenamefont {Kemper}, \citenamefont {McDaniel}, \citenamefont {Obertelli}, \citenamefont {Otsuka}, \citenamefont {Ratkiewicz}, \citenamefont {Terry}, \citenamefont {Utsuno},\ and\ \citenamefont {Weisshaar}}]{Gade2011}%
  \BibitemOpen
  \bibfield  {author} {\bibinfo {author} {\bibfnamefont {A.}~\bibnamefont {Gade}}, \bibinfo {author} {\bibfnamefont {D.}~\bibnamefont {Bazin}}, \bibinfo {author} {\bibfnamefont {B.~A.}\ \bibnamefont {Brown}}, \bibinfo {author} {\bibfnamefont {C.~M.}\ \bibnamefont {Campbell}}, \bibinfo {author} {\bibfnamefont {J.~M.}\ \bibnamefont {Cook}}, \bibinfo {author} {\bibfnamefont {S.}~\bibnamefont {Ettenauer}}, \bibinfo {author} {\bibfnamefont {T.}~\bibnamefont {Glasmacher}}, \bibinfo {author} {\bibfnamefont {K.~W.}\ \bibnamefont {Kemper}}, \bibinfo {author} {\bibfnamefont {S.}~\bibnamefont {McDaniel}}, \bibinfo {author} {\bibfnamefont {A.}~\bibnamefont {Obertelli}}, \bibinfo {author} {\bibfnamefont {T.}~\bibnamefont {Otsuka}}, \bibinfo {author} {\bibfnamefont {A.}~\bibnamefont {Ratkiewicz}}, \bibinfo {author} {\bibfnamefont {J.~R.}\ \bibnamefont {Terry}}, \bibinfo {author} {\bibfnamefont {Y.}~\bibnamefont {Utsuno}}, \ and\ \bibinfo {author} {\bibfnamefont {D.}~\bibnamefont {Weisshaar}},\ }\href {\doibase
  10.1103/PhysRevC.83.044305} {\bibfield  {journal} {\bibinfo  {journal} {Phys. Rev. C}\ }\textbf {\bibinfo {volume} {83}},\ \bibinfo {pages} {044305} (\bibinfo {year} {2011})}\BibitemShut {NoStop}%
\bibitem [{\citenamefont {Gaudefroy}\ \emph {et~al.}(2012)\citenamefont {Gaudefroy}, \citenamefont {Mittig}, \citenamefont {Orr}, \citenamefont {Varet}, \citenamefont {Chartier}, \citenamefont {Roussel-Chomaz}, \citenamefont {Ebran}, \citenamefont {Fern\'andez-Dom\'{\i}nguez}, \citenamefont {Fr\'emont}, \citenamefont {Gangnant}, \citenamefont {Gillibert}, \citenamefont {Gr\'evy}, \citenamefont {Libin}, \citenamefont {Maslov}, \citenamefont {Paschalis}, \citenamefont {Pietras}, \citenamefont {Penionzhkevich}, \citenamefont {Spitaels},\ and\ \citenamefont {Villari}}]{Gaudfroy2012}%
  \BibitemOpen
  \bibfield  {author} {\bibinfo {author} {\bibfnamefont {L.}~\bibnamefont {Gaudefroy}}, \bibinfo {author} {\bibfnamefont {W.}~\bibnamefont {Mittig}}, \bibinfo {author} {\bibfnamefont {N.~A.}\ \bibnamefont {Orr}}, \bibinfo {author} {\bibfnamefont {S.}~\bibnamefont {Varet}}, \bibinfo {author} {\bibfnamefont {M.}~\bibnamefont {Chartier}}, \bibinfo {author} {\bibfnamefont {P.}~\bibnamefont {Roussel-Chomaz}}, \bibinfo {author} {\bibfnamefont {J.~P.}\ \bibnamefont {Ebran}}, \bibinfo {author} {\bibfnamefont {B.}~\bibnamefont {Fern\'andez-Dom\'{\i}nguez}}, \bibinfo {author} {\bibfnamefont {G.}~\bibnamefont {Fr\'emont}}, \bibinfo {author} {\bibfnamefont {P.}~\bibnamefont {Gangnant}}, \bibinfo {author} {\bibfnamefont {A.}~\bibnamefont {Gillibert}}, \bibinfo {author} {\bibfnamefont {S.}~\bibnamefont {Gr\'evy}}, \bibinfo {author} {\bibfnamefont {J.~F.}\ \bibnamefont {Libin}}, \bibinfo {author} {\bibfnamefont {V.~A.}\ \bibnamefont {Maslov}}, \bibinfo {author} {\bibfnamefont {S.}~\bibnamefont {Paschalis}}, \bibinfo
  {author} {\bibfnamefont {B.}~\bibnamefont {Pietras}}, \bibinfo {author} {\bibfnamefont {Y.-E.}\ \bibnamefont {Penionzhkevich}}, \bibinfo {author} {\bibfnamefont {C.}~\bibnamefont {Spitaels}}, \ and\ \bibinfo {author} {\bibfnamefont {A.~C.~C.}\ \bibnamefont {Villari}},\ }\href {\doibase 10.1103/PhysRevLett.109.202503} {\bibfield  {journal} {\bibinfo  {journal} {Phys. Rev. Lett.}\ }\textbf {\bibinfo {volume} {109}},\ \bibinfo {pages} {202503} (\bibinfo {year} {2012})}\BibitemShut {NoStop}%
\bibitem [{\citenamefont {Kuboki}\ \emph {et~al.}(2011)\citenamefont {Kuboki}, \citenamefont {Ohtsubo}, \citenamefont {Takechi}, \citenamefont {Hachiuma},\ and\ \citenamefont {Namihira}}]{Kuboki2011}%
  \BibitemOpen
  \bibfield  {author} {\bibinfo {author} {\bibfnamefont {T.}~\bibnamefont {Kuboki}}, \bibinfo {author} {\bibfnamefont {T.}~\bibnamefont {Ohtsubo}}, \bibinfo {author} {\bibfnamefont {M.}~\bibnamefont {Takechi}}, \bibinfo {author} {\bibfnamefont {I.}~\bibnamefont {Hachiuma}}, \ and\ \bibinfo {author} {\bibfnamefont {K.}~\bibnamefont {Namihira}},\ }\href {\doibase 10.5506/aphyspolb.42.765} {\bibfield  {journal} {\bibinfo  {journal} {Acta Physica Polonica B}\ }\textbf {\bibinfo {volume} {42}},\ \bibinfo {pages} {765} (\bibinfo {year} {2011})}\BibitemShut {NoStop}%
\bibitem [{\citenamefont {Suzuki}\ \emph {et~al.}(2014)\citenamefont {Suzuki}, \citenamefont {Takechi}, \citenamefont {Ohtsubo}, \citenamefont {Nishimura}, \citenamefont {Fukuda}, \citenamefont {Kuboki}, \citenamefont {Nagashima}, \citenamefont {Suzuki}, \citenamefont {Yamaguchi}, \citenamefont {Ozawa}, \citenamefont {Ohishi}, \citenamefont {Moriguchi}, \citenamefont {Sumikama}, \citenamefont {Geissel}, \citenamefont {Aoi}, \citenamefont {Chen}, \citenamefont {Fang}, \citenamefont {Fukuda}, \citenamefont {Fukuoka}, \citenamefont {Furuki}, \citenamefont {Inabe}, \citenamefont {Ishibashi}, \citenamefont {Ito}, \citenamefont {Izumikawa}, \citenamefont {Kameda}, \citenamefont {Ueda}, \citenamefont {Lantz}, \citenamefont {Lee}, \citenamefont {G.}, \citenamefont {Mihara}, \citenamefont {Momota}, \citenamefont {Nagae}, \citenamefont {Nishikiori}, \citenamefont {Niwa}, \citenamefont {Ohnishi}, \citenamefont {Okumura}, \citenamefont {Ogura}, \citenamefont {Sakuraï}, \citenamefont {Sato}, \citenamefont {Shimbara},
  \citenamefont {Suzuki}, \citenamefont {Takeda}, \citenamefont {Takeuchi}, \citenamefont {Tanaka}, \citenamefont {Uenishi}, \citenamefont {Winkler},\ and\ \citenamefont {Yanagisawa}}]{Suzuki2014}%
  \BibitemOpen
  \bibfield  {author} {\bibinfo {author} {\bibfnamefont {S.}~\bibnamefont {Suzuki}}, \bibinfo {author} {\bibfnamefont {M.}~\bibnamefont {Takechi}}, \bibinfo {author} {\bibfnamefont {T.}~\bibnamefont {Ohtsubo}}, \bibinfo {author} {\bibfnamefont {D.}~\bibnamefont {Nishimura}}, \bibinfo {author} {\bibfnamefont {M.}~\bibnamefont {Fukuda}}, \bibinfo {author} {\bibfnamefont {T.}~\bibnamefont {Kuboki}}, \bibinfo {author} {\bibfnamefont {M.}~\bibnamefont {Nagashima}}, \bibinfo {author} {\bibfnamefont {T.}~\bibnamefont {Suzuki}}, \bibinfo {author} {\bibfnamefont {T.}~\bibnamefont {Yamaguchi}}, \bibinfo {author} {\bibfnamefont {A.}~\bibnamefont {Ozawa}}, \bibinfo {author} {\bibfnamefont {H.}~\bibnamefont {Ohishi}}, \bibinfo {author} {\bibfnamefont {T.}~\bibnamefont {Moriguchi}}, \bibinfo {author} {\bibfnamefont {T.}~\bibnamefont {Sumikama}}, \bibinfo {author} {\bibfnamefont {H.}~\bibnamefont {Geissel}}, \bibinfo {author} {\bibfnamefont {N.}~\bibnamefont {Aoi}}, \bibinfo {author} {\bibfnamefont {R.}~\bibnamefont
  {Chen}}, \bibinfo {author} {\bibfnamefont {D.}~\bibnamefont {Fang}}, \bibinfo {author} {\bibfnamefont {N.}~\bibnamefont {Fukuda}}, \bibinfo {author} {\bibfnamefont {S.}~\bibnamefont {Fukuoka}}, \bibinfo {author} {\bibfnamefont {H.}~\bibnamefont {Furuki}}, \bibinfo {author} {\bibfnamefont {N.}~\bibnamefont {Inabe}}, \bibinfo {author} {\bibfnamefont {Y.}~\bibnamefont {Ishibashi}}, \bibinfo {author} {\bibfnamefont {T.}~\bibnamefont {Ito}}, \bibinfo {author} {\bibfnamefont {T.}~\bibnamefont {Izumikawa}}, \bibinfo {author} {\bibfnamefont {D.}~\bibnamefont {Kameda}}, \bibinfo {author} {\bibfnamefont {T.}~\bibnamefont {Ueda}}, \bibinfo {author} {\bibfnamefont {M.}~\bibnamefont {Lantz}}, \bibinfo {author} {\bibfnamefont {C.~S.}\ \bibnamefont {Lee}}, \bibinfo {author} {\bibfnamefont {Y.}~\bibnamefont {G.}}, \bibinfo {author} {\bibfnamefont {M.}~\bibnamefont {Mihara}}, \bibinfo {author} {\bibfnamefont {S.}~\bibnamefont {Momota}}, \bibinfo {author} {\bibfnamefont {D.}~\bibnamefont {Nagae}}, \bibinfo {author}
  {\bibfnamefont {R.}~\bibnamefont {Nishikiori}}, \bibinfo {author} {\bibfnamefont {T.}~\bibnamefont {Niwa}}, \bibinfo {author} {\bibfnamefont {T.}~\bibnamefont {Ohnishi}}, \bibinfo {author} {\bibfnamefont {K.}~\bibnamefont {Okumura}}, \bibinfo {author} {\bibfnamefont {T.}~\bibnamefont {Ogura}}, \bibinfo {author} {\bibfnamefont {H.}~\bibnamefont {Sakuraï}}, \bibinfo {author} {\bibfnamefont {K.}~\bibnamefont {Sato}}, \bibinfo {author} {\bibfnamefont {Y.}~\bibnamefont {Shimbara}}, \bibinfo {author} {\bibfnamefont {H.}~\bibnamefont {Suzuki}}, \bibinfo {author} {\bibfnamefont {H.}~\bibnamefont {Takeda}}, \bibinfo {author} {\bibfnamefont {S.}~\bibnamefont {Takeuchi}}, \bibinfo {author} {\bibfnamefont {K.}~\bibnamefont {Tanaka}}, \bibinfo {author} {\bibfnamefont {H.}~\bibnamefont {Uenishi}}, \bibinfo {author} {\bibfnamefont {M.}~\bibnamefont {Winkler}}, \ and\ \bibinfo {author} {\bibfnamefont {Y.}~\bibnamefont {Yanagisawa}},\ }\href {\doibase 10.1051/epjconf/20146603084} {\bibfield  {journal} {\bibinfo  {journal}
  {EPJ Web of Conferences}\ }\textbf {\bibinfo {volume} {66}},\ \bibinfo {pages} {03084} (\bibinfo {year} {2014})}\BibitemShut {NoStop}%
\bibitem [{\citenamefont {Fortune}\ and\ \citenamefont {Sherr}(2013)}]{Fortune2013}%
  \BibitemOpen
  \bibfield  {author} {\bibinfo {author} {\bibfnamefont {H.~T.}\ \bibnamefont {Fortune}}\ and\ \bibinfo {author} {\bibfnamefont {R.}~\bibnamefont {Sherr}},\ }\href {\doibase 10.1103/PhysRevC.87.057308} {\bibfield  {journal} {\bibinfo  {journal} {Phys. Rev. C}\ }\textbf {\bibinfo {volume} {87}},\ \bibinfo {pages} {057308} (\bibinfo {year} {2013})}\BibitemShut {NoStop}%
\bibitem [{\citenamefont {Singh}\ \emph {et~al.}(2016)\citenamefont {Singh}, \citenamefont {Shubhchintak},\ and\ \citenamefont {Chatterjee}}]{GSingh2016}%
  \BibitemOpen
  \bibfield  {author} {\bibinfo {author} {\bibfnamefont {G.}~\bibnamefont {Singh}}, \bibinfo {author} {\bibnamefont {Shubhchintak}}, \ and\ \bibinfo {author} {\bibfnamefont {R.}~\bibnamefont {Chatterjee}},\ }\href {\doibase 10.1103/PhysRevC.94.024606} {\bibfield  {journal} {\bibinfo  {journal} {Phys. Rev. C}\ }\textbf {\bibinfo {volume} {94}},\ \bibinfo {pages} {024606} (\bibinfo {year} {2016})}\BibitemShut {NoStop}%
\bibitem [{\citenamefont {Singh}\ \emph {et~al.}(2017)\citenamefont {Singh}, \citenamefont {Shubhchintak},\ and\ \citenamefont {Chatterjee}}]{GSingh2017}%
  \BibitemOpen
  \bibfield  {author} {\bibinfo {author} {\bibfnamefont {G.}~\bibnamefont {Singh}}, \bibinfo {author} {\bibnamefont {Shubhchintak}}, \ and\ \bibinfo {author} {\bibfnamefont {R.}~\bibnamefont {Chatterjee}},\ }\href {\doibase 10.1103/PhysRevC.95.065806} {\bibfield  {journal} {\bibinfo  {journal} {Phys. Rev. C}\ }\textbf {\bibinfo {volume} {95}},\ \bibinfo {pages} {065806} (\bibinfo {year} {2017})}\BibitemShut {NoStop}%
\bibitem [{\citenamefont {Manju}\ \emph {et~al.}(2019)\citenamefont {Manju}, \citenamefont {Singh}, \citenamefont {Shubhchintak},\ and\ \citenamefont {Chatterjee}}]{Manju19EPJ}%
  \BibitemOpen
  \bibfield  {author} {\bibinfo {author} {\bibnamefont {Manju}}, \bibinfo {author} {\bibfnamefont {J.}~\bibnamefont {Singh}}, \bibinfo {author} {\bibnamefont {Shubhchintak}}, \ and\ \bibinfo {author} {\bibfnamefont {R.}~\bibnamefont {Chatterjee}},\ }\href {\doibase 10.1140/epja/i2019-12679-4} {\bibfield  {journal} {\bibinfo  {journal} {The European Physical Journal A}\ }\textbf {\bibinfo {volume} {55}},\ \bibinfo {pages} {5} (\bibinfo {year} {2019})}\BibitemShut {NoStop}%
\bibitem [{\citenamefont {Luo}\ \emph {et~al.}(2020)\citenamefont {Luo}, \citenamefont {Liu}, \citenamefont {Guo},\ and\ \citenamefont {Yang}}]{Luo2020}%
  \BibitemOpen
  \bibfield  {author} {\bibinfo {author} {\bibfnamefont {Y.}~\bibnamefont {Luo}}, \bibinfo {author} {\bibfnamefont {Q.}~\bibnamefont {Liu}}, \bibinfo {author} {\bibfnamefont {J.}~\bibnamefont {Guo}}, \ and\ \bibinfo {author} {\bibfnamefont {Y.~Y.}\ \bibnamefont {Yang}},\ }\href {\doibase 10.1088/1361-6471/ab92e2} {\bibfield  {journal} {\bibinfo  {journal} {Journal of Physics G: Nuclear and Particle Physics}\ }\textbf {\bibinfo {volume} {47}},\ \bibinfo {pages} {085105} (\bibinfo {year} {2020})}\BibitemShut {NoStop}%
\bibitem [{\citenamefont {Ahn}\ \emph {et~al.}(2022)\citenamefont {Ahn}, \citenamefont {Amano}, \citenamefont {Baba}, \citenamefont {Fukuda}, \citenamefont {Geissel}, \citenamefont {Inabe}, \citenamefont {Ishikawa}, \citenamefont {Iwasa}, \citenamefont {Komatsubara}, \citenamefont {Kubo}, \citenamefont {Kusaka}, \citenamefont {Morrissey}, \citenamefont {Nakamura}, \citenamefont {Ohtake}, \citenamefont {Otsu}, \citenamefont {Sakakibara}, \citenamefont {Sato}, \citenamefont {Sherrill}, \citenamefont {Shimizu}, \citenamefont {Sumikama}, \citenamefont {Suzuki}, \citenamefont {Takeda}, \citenamefont {Tarasov}, \citenamefont {Ueno}, \citenamefont {Yanagisawa},\ and\ \citenamefont {Yoshida}}]{Ahn2022}%
  \BibitemOpen
  \bibfield  {author} {\bibinfo {author} {\bibfnamefont {D.~S.}\ \bibnamefont {Ahn}}, \bibinfo {author} {\bibfnamefont {J.}~\bibnamefont {Amano}}, \bibinfo {author} {\bibfnamefont {H.}~\bibnamefont {Baba}}, \bibinfo {author} {\bibfnamefont {N.}~\bibnamefont {Fukuda}}, \bibinfo {author} {\bibfnamefont {H.}~\bibnamefont {Geissel}}, \bibinfo {author} {\bibfnamefont {N.}~\bibnamefont {Inabe}}, \bibinfo {author} {\bibfnamefont {S.}~\bibnamefont {Ishikawa}}, \bibinfo {author} {\bibfnamefont {N.}~\bibnamefont {Iwasa}}, \bibinfo {author} {\bibfnamefont {T.}~\bibnamefont {Komatsubara}}, \bibinfo {author} {\bibfnamefont {T.}~\bibnamefont {Kubo}}, \bibinfo {author} {\bibfnamefont {K.}~\bibnamefont {Kusaka}}, \bibinfo {author} {\bibfnamefont {D.~J.}\ \bibnamefont {Morrissey}}, \bibinfo {author} {\bibfnamefont {T.}~\bibnamefont {Nakamura}}, \bibinfo {author} {\bibfnamefont {M.}~\bibnamefont {Ohtake}}, \bibinfo {author} {\bibfnamefont {H.}~\bibnamefont {Otsu}}, \bibinfo {author} {\bibfnamefont {T.}~\bibnamefont
  {Sakakibara}}, \bibinfo {author} {\bibfnamefont {H.}~\bibnamefont {Sato}}, \bibinfo {author} {\bibfnamefont {B.~M.}\ \bibnamefont {Sherrill}}, \bibinfo {author} {\bibfnamefont {Y.}~\bibnamefont {Shimizu}}, \bibinfo {author} {\bibfnamefont {T.}~\bibnamefont {Sumikama}}, \bibinfo {author} {\bibfnamefont {H.}~\bibnamefont {Suzuki}}, \bibinfo {author} {\bibfnamefont {H.}~\bibnamefont {Takeda}}, \bibinfo {author} {\bibfnamefont {O.~B.}\ \bibnamefont {Tarasov}}, \bibinfo {author} {\bibfnamefont {H.}~\bibnamefont {Ueno}}, \bibinfo {author} {\bibfnamefont {Y.}~\bibnamefont {Yanagisawa}}, \ and\ \bibinfo {author} {\bibfnamefont {K.}~\bibnamefont {Yoshida}},\ }\href {\doibase 10.1103/PhysRevLett.129.212502} {\bibfield  {journal} {\bibinfo  {journal} {Phys. Rev. Lett.}\ }\textbf {\bibinfo {volume} {129}},\ \bibinfo {pages} {212502} (\bibinfo {year} {2022})}\BibitemShut {NoStop}%
\bibitem [{\citenamefont {Zhang}\ \emph {et~al.}(2023)\citenamefont {Zhang}, \citenamefont {Papakonstantinou}, \citenamefont {Mun}, \citenamefont {Kim}, \citenamefont {Yan},\ and\ \citenamefont {Sun}}]{Zhang2023}%
  \BibitemOpen
  \bibfield  {author} {\bibinfo {author} {\bibfnamefont {K.~Y.}\ \bibnamefont {Zhang}}, \bibinfo {author} {\bibfnamefont {P.}~\bibnamefont {Papakonstantinou}}, \bibinfo {author} {\bibfnamefont {M.-H.}\ \bibnamefont {Mun}}, \bibinfo {author} {\bibfnamefont {Y.}~\bibnamefont {Kim}}, \bibinfo {author} {\bibfnamefont {H.}~\bibnamefont {Yan}}, \ and\ \bibinfo {author} {\bibfnamefont {X.-X.}\ \bibnamefont {Sun}},\ }\href {\doibase 10.1103/PhysRevC.107.L041303} {\bibfield  {journal} {\bibinfo  {journal} {Phys. Rev. C}\ }\textbf {\bibinfo {volume} {107}},\ \bibinfo {pages} {L041303} (\bibinfo {year} {2023})}\BibitemShut {NoStop}%
\bibitem [{\citenamefont {Lay}\ \emph {et~al.}(2010)\citenamefont {Lay}, \citenamefont {Moro}, \citenamefont {Arias},\ and\ \citenamefont {G\'omez-Camacho}}]{JALay10}%
  \BibitemOpen
  \bibfield  {author} {\bibinfo {author} {\bibfnamefont {J.~A.}\ \bibnamefont {Lay}}, \bibinfo {author} {\bibfnamefont {A.~M.}\ \bibnamefont {Moro}}, \bibinfo {author} {\bibfnamefont {J.~M.}\ \bibnamefont {Arias}}, \ and\ \bibinfo {author} {\bibfnamefont {J.}~\bibnamefont {G\'omez-Camacho}},\ }\href {\doibase 10.1103/PhysRevC.82.024605} {\bibfield  {journal} {\bibinfo  {journal} {Phys. Rev. C}\ }\textbf {\bibinfo {volume} {82}},\ \bibinfo {pages} {024605} (\bibinfo {year} {2010})}\BibitemShut {NoStop}%
\bibitem [{\citenamefont {Casal}\ \emph {et~al.}(2013)\citenamefont {Casal}, \citenamefont {Rodr\'{\i}guez-Gallardo},\ and\ \citenamefont {Arias}}]{JCasal13}%
  \BibitemOpen
  \bibfield  {author} {\bibinfo {author} {\bibfnamefont {J.}~\bibnamefont {Casal}}, \bibinfo {author} {\bibfnamefont {M.}~\bibnamefont {Rodr\'{\i}guez-Gallardo}}, \ and\ \bibinfo {author} {\bibfnamefont {J.~M.}\ \bibnamefont {Arias}},\ }\href {\doibase 10.1103/PhysRevC.88.014327} {\bibfield  {journal} {\bibinfo  {journal} {Phys. Rev. C}\ }\textbf {\bibinfo {volume} {88}},\ \bibinfo {pages} {014327} (\bibinfo {year} {2013})}\BibitemShut {NoStop}%
\bibitem [{\citenamefont {Zhukov}\ \emph {et~al.}(1993)\citenamefont {Zhukov}, \citenamefont {Danilin}, \citenamefont {Fedorov}, \citenamefont {Bang}, \citenamefont {Thompson},\ and\ \citenamefont {Vaagen}}]{ZHUKOV1993}%
  \BibitemOpen
  \bibfield  {author} {\bibinfo {author} {\bibfnamefont {M.}~\bibnamefont {Zhukov}}, \bibinfo {author} {\bibfnamefont {B.}~\bibnamefont {Danilin}}, \bibinfo {author} {\bibfnamefont {D.}~\bibnamefont {Fedorov}}, \bibinfo {author} {\bibfnamefont {J.}~\bibnamefont {Bang}}, \bibinfo {author} {\bibfnamefont {I.}~\bibnamefont {Thompson}}, \ and\ \bibinfo {author} {\bibfnamefont {J.}~\bibnamefont {Vaagen}},\ }\href {\doibase https://doi.org/10.1016/0370-1573(93)90141-Y} {\bibfield  {journal} {\bibinfo  {journal} {Physics Reports}\ }\textbf {\bibinfo {volume} {231}},\ \bibinfo {pages} {151} (\bibinfo {year} {1993})}\BibitemShut {NoStop}%
\bibitem [{\citenamefont {Nielsen}\ \emph {et~al.}(2001)\citenamefont {Nielsen}, \citenamefont {Fedorov}, \citenamefont {Jensen},\ and\ \citenamefont {Garrido}}]{Nielsen01}%
  \BibitemOpen
  \bibfield  {author} {\bibinfo {author} {\bibfnamefont {E.}~\bibnamefont {Nielsen}}, \bibinfo {author} {\bibfnamefont {D.}~\bibnamefont {Fedorov}}, \bibinfo {author} {\bibfnamefont {A.}~\bibnamefont {Jensen}}, \ and\ \bibinfo {author} {\bibfnamefont {E.}~\bibnamefont {Garrido}},\ }\href {\doibase https://doi.org/10.1016/S0370-1573(00)00107-1} {\bibfield  {journal} {\bibinfo  {journal} {Physics Reports}\ }\textbf {\bibinfo {volume} {347}},\ \bibinfo {pages} {373 } (\bibinfo {year} {2001})}\BibitemShut {NoStop}%
\bibitem [{\citenamefont {Casal}\ \emph {et~al.}(2020)\citenamefont {Casal}, \citenamefont {Singh}, \citenamefont {Fortunato}, \citenamefont {Horiuchi},\ and\ \citenamefont {Vitturi}}]{JCasal20}%
  \BibitemOpen
  \bibfield  {author} {\bibinfo {author} {\bibfnamefont {J.}~\bibnamefont {Casal}}, \bibinfo {author} {\bibfnamefont {J.}~\bibnamefont {Singh}}, \bibinfo {author} {\bibfnamefont {L.}~\bibnamefont {Fortunato}}, \bibinfo {author} {\bibfnamefont {W.}~\bibnamefont {Horiuchi}}, \ and\ \bibinfo {author} {\bibfnamefont {A.}~\bibnamefont {Vitturi}},\ }\href {\doibase 10.1103/PhysRevC.102.064627} {\bibfield  {journal} {\bibinfo  {journal} {Phys. Rev. C}\ }\textbf {\bibinfo {volume} {102}},\ \bibinfo {pages} {064627} (\bibinfo {year} {2020})}\BibitemShut {NoStop}%
\bibitem [{\citenamefont {Horiuchi}\ \emph {et~al.}(2010)\citenamefont {Horiuchi}, \citenamefont {Suzuki}, \citenamefont {Capel},\ and\ \citenamefont {Baye}}]{Horiuchi201031Ne}%
  \BibitemOpen
  \bibfield  {author} {\bibinfo {author} {\bibfnamefont {W.}~\bibnamefont {Horiuchi}}, \bibinfo {author} {\bibfnamefont {Y.}~\bibnamefont {Suzuki}}, \bibinfo {author} {\bibfnamefont {P.}~\bibnamefont {Capel}}, \ and\ \bibinfo {author} {\bibfnamefont {D.}~\bibnamefont {Baye}},\ }\href {\doibase 10.1103/PhysRevC.81.024606} {\bibfield  {journal} {\bibinfo  {journal} {Phys. Rev. C}\ }\textbf {\bibinfo {volume} {81}},\ \bibinfo {pages} {024606} (\bibinfo {year} {2010})}\BibitemShut {NoStop}%
\bibitem [{\citenamefont {Singh}\ \emph {et~al.}(2020)\citenamefont {Singh}, \citenamefont {Casal}, \citenamefont {Horiuchi}, \citenamefont {Fortunato},\ and\ \citenamefont {Vitturi}}]{JSingh2020}%
  \BibitemOpen
  \bibfield  {author} {\bibinfo {author} {\bibfnamefont {J.}~\bibnamefont {Singh}}, \bibinfo {author} {\bibfnamefont {J.}~\bibnamefont {Casal}}, \bibinfo {author} {\bibfnamefont {W.}~\bibnamefont {Horiuchi}}, \bibinfo {author} {\bibfnamefont {L.}~\bibnamefont {Fortunato}}, \ and\ \bibinfo {author} {\bibfnamefont {A.}~\bibnamefont {Vitturi}},\ }\href {\doibase 10.1103/PhysRevC.101.024310} {\bibfield  {journal} {\bibinfo  {journal} {Phys. Rev. C}\ }\textbf {\bibinfo {volume} {101}},\ \bibinfo {pages} {024310} (\bibinfo {year} {2020})}\BibitemShut {NoStop}%
\bibitem [{\citenamefont {Singh}\ \emph {et~al.}(2022)\citenamefont {Singh}, \citenamefont {Singh}, \citenamefont {Casal},\ and\ \citenamefont {Fortunato}}]{GSingh2022}%
  \BibitemOpen
  \bibfield  {author} {\bibinfo {author} {\bibfnamefont {G.}~\bibnamefont {Singh}}, \bibinfo {author} {\bibfnamefont {J.}~\bibnamefont {Singh}}, \bibinfo {author} {\bibfnamefont {J.}~\bibnamefont {Casal}}, \ and\ \bibinfo {author} {\bibfnamefont {L.}~\bibnamefont {Fortunato}},\ }\href {\doibase 10.1103/PhysRevC.105.014328} {\bibfield  {journal} {\bibinfo  {journal} {Phys. Rev. C}\ }\textbf {\bibinfo {volume} {105}},\ \bibinfo {pages} {014328} (\bibinfo {year} {2022})}\BibitemShut {NoStop}%
\bibitem [{\citenamefont {Geng}\ \emph {et~al.}(2004)\citenamefont {Geng}, \citenamefont {Toki}, \citenamefont {Ozawa},\ and\ \citenamefont {Meng}}]{GENG2004}%
  \BibitemOpen
  \bibfield  {author} {\bibinfo {author} {\bibfnamefont {L.}~\bibnamefont {Geng}}, \bibinfo {author} {\bibfnamefont {H.}~\bibnamefont {Toki}}, \bibinfo {author} {\bibfnamefont {A.}~\bibnamefont {Ozawa}}, \ and\ \bibinfo {author} {\bibfnamefont {J.}~\bibnamefont {Meng}},\ }\href {\doibase https://doi.org/10.1016/j.nuclphysa.2003.10.014} {\bibfield  {journal} {\bibinfo  {journal} {Nuclear Physics A}\ }\textbf {\bibinfo {volume} {730}},\ \bibinfo {pages} {80} (\bibinfo {year} {2004})}\BibitemShut {NoStop}%
\bibitem [{\citenamefont {Otsuka}\ \emph {et~al.}(2022)\citenamefont {Otsuka}, \citenamefont {Shimizu},\ and\ \citenamefont {Tsunoda}}]{Otsuka2022}%
  \BibitemOpen
  \bibfield  {author} {\bibinfo {author} {\bibfnamefont {T.}~\bibnamefont {Otsuka}}, \bibinfo {author} {\bibfnamefont {N.}~\bibnamefont {Shimizu}}, \ and\ \bibinfo {author} {\bibfnamefont {Y.}~\bibnamefont {Tsunoda}},\ }\href {\doibase 10.1103/PhysRevC.105.014319} {\bibfield  {journal} {\bibinfo  {journal} {Phys. Rev. C}\ }\textbf {\bibinfo {volume} {105}},\ \bibinfo {pages} {014319} (\bibinfo {year} {2022})}\BibitemShut {NoStop}%
\bibitem [{\citenamefont {Gogny}\ \emph {et~al.}(1970)\citenamefont {Gogny}, \citenamefont {Pires},\ and\ \citenamefont {{De Tourreil}}}]{Gogny70}%
  \BibitemOpen
  \bibfield  {author} {\bibinfo {author} {\bibfnamefont {D.}~\bibnamefont {Gogny}}, \bibinfo {author} {\bibfnamefont {P.}~\bibnamefont {Pires}}, \ and\ \bibinfo {author} {\bibfnamefont {R.}~\bibnamefont {{De Tourreil}}},\ }\href {\doibase https://doi.org/10.1016/0370-2693(70)90552-6} {\bibfield  {journal} {\bibinfo  {journal} {Phys. Lett. B}\ }\textbf {\bibinfo {volume} {32}},\ \bibinfo {pages} {591} (\bibinfo {year} {1970})}\BibitemShut {NoStop}%
\bibitem [{\citenamefont {Singh}\ \emph {et~al.}(2024)\citenamefont {Singh}, \citenamefont {Casal}, \citenamefont {Horiuchi}, \citenamefont {Walet},\ and\ \citenamefont {Satuła}}]{JSingh2024}%
  \BibitemOpen
  \bibfield  {author} {\bibinfo {author} {\bibfnamefont {J.}~\bibnamefont {Singh}}, \bibinfo {author} {\bibfnamefont {J.}~\bibnamefont {Casal}}, \bibinfo {author} {\bibfnamefont {W.}~\bibnamefont {Horiuchi}}, \bibinfo {author} {\bibfnamefont {N.~R.}\ \bibnamefont {Walet}}, \ and\ \bibinfo {author} {\bibfnamefont {W.}~\bibnamefont {Satuła}},\ }\href {\doibase https://doi.org/10.1016/j.physletb.2024.138694} {\bibfield  {journal} {\bibinfo  {journal} {Phys. Lett. B}\ }\textbf {\bibinfo {volume} {853}},\ \bibinfo {pages} {138694} (\bibinfo {year} {2024})}\BibitemShut {NoStop}%
\bibitem [{\citenamefont {Moro}\ and\ \citenamefont {Lay}(2012)}]{Amoro12}%
  \BibitemOpen
  \bibfield  {author} {\bibinfo {author} {\bibfnamefont {A.~M.}\ \bibnamefont {Moro}}\ and\ \bibinfo {author} {\bibfnamefont {J.~A.}\ \bibnamefont {Lay}},\ }\href {\doibase 10.1103/PhysRevLett.109.232502} {\bibfield  {journal} {\bibinfo  {journal} {Phys. Rev. Lett.}\ }\textbf {\bibinfo {volume} {109}},\ \bibinfo {pages} {232502} (\bibinfo {year} {2012})}\BibitemShut {NoStop}%
\end{thebibliography}%
\end{document}